\documentclass[sn-apa]{sn-jnl}% APA Reference Style
\usepackage{appendix}
\usepackage{multirow}
\usepackage{graphicx}
\usepackage{amssymb}
\usepackage{csquotes}
\usepackage[inline]{enumitem}
\usepackage{amsmath}
\usepackage{xcolor}
\usepackage{hyperref}
\usepackage{mathbbol}
\usepackage{booktabs}
\usepackage[usestackEOL]{stackengine}

\DeclareMathOperator*{\argmax}{arg\,max}

\usepackage{caption}
\usepackage{float}
\jyear{2021}%

\theoremstyle{thmstyleone}%
\theoremstyle{thmstyletwo}%

\theoremstyle{thmstylethree}%

\begin{document}

\title[Interactions between dynamic team composition and coordination]{Interactions between dynamic team composition and coordination}
\subtitle{An agent-based modeling approach}


\author*[1]{\fnm{Darío} \sur{Blanco-Fernández}}\email{dario.blanco@aau.at}
\equalcont{These authors contributed equally to this work.}

\author[2]{\fnm{Stephan} \sur{Leitner}}\email{stephan.leitner@aau.at}
\equalcont{These authors contributed equally to this work.}

\author[2]{\fnm{Alexandra} \sur{Rausch}}\email{alexandra.rausch@aau.at}
\equalcont{These authors contributed equally to this work.}

\affil[1]{\orgdiv{Digital Age Research Center}, \orgname{University of Klagenfurt}, \orgaddress{\street{Universitätsstraße 65-67}, \city{Klagenfurt}, \postcode{9020}, \country{Austria}}}

\affil[2]{\orgdiv{Department of Management Control and Strategic Management}, \orgname{University of Klagenfurt}, \orgaddress{\street{Universitätsstraße 65-67}, \city{Klagenfurt}, \postcode{9020}, \country{Austria}}}


\abstract{This paper examines the interactions between selected coordination modes and dynamic team composition, and their joint effects on task performance under different task complexity and individual learning conditions. Prior research often treats dynamic team composition as a consequence of suboptimal organizational design choices. The emergence of new organizational forms that consciously employ teams that change their composition periodically challenges this perspective. In this paper, we follow the contingency theory and characterize dynamic team composition as a design choice that interacts with other choices such as the coordination mode, and with additional contextual factors such as individual learning and task complexity. We employ an agent-based modeling approach based on the NK framework, which includes a reinforcement learning mechanism, a recurring team formation mechanism based on signaling, and three different coordination modes. Our results suggest that by implementing lateral communication or sequential decision-making, teams may exploit the benefits of dynamic composition more than if decision-making is fully autonomous. The choice of a proper coordination mode, however, is partly moderated by the task complexity and individual learning. Additionally, we show that only a coordination mode based on lateral communication may prevent the negative effects of individual learning.}




\maketitle

\newpage

\section{Introduction}
The traditional perspective on dynamic team composition---i.e., the periodic substitution of team members with different individuals---characterizes it as a negative outcome of suboptimal organizational design choices \citep{Carley1992,Davis1973}. This perspective connects dynamic team composition to the concept of organizational turnover, leading to the common assumption that dynamic team composition results in detrimental effects for organizations \citep{Tannenbaum2012}. Specifically, prior research \citep{Edmondsonetal2003,Gardner2010,Harrison2003} argues that changing the composition of a team disrupts the development of autonomous coordination among team members---meaning the flow and exchange of information between individuals regarding their potential actions \citep{Burton2005,Galbraith1973,Nadler1997,Leitner2020}. This reduced coordination, in turn, leads to a decline in task performance---i.e., in the effectiveness with which a task is completed \citep{Carley1992,Lin1992,Littlepage1997,Reagans2005}.

The emergence of new organizational forms in recent years, however, challenges this traditional understanding of dynamic team composition. During the past decades, many countries have experienced a process of transition from an industrial towards a knowledge-based economy \citep{Lin2023}. Furthermore, these last decades have witnessed the unstoppable advance of digitalization, impacting almost every facet of our everyday lives \citep{Calderon-Monge2023}. Consequently, organizations and the environment in which they operate have changed substantially. Since the tasks and problems faced by organizations have become more complex, teams experience higher uncertainty when handling them \citep{Rojas-Cordova2022}. This increased complexity and uncertainty require teams to be more flexible, adaptative and agile \citep{Burton2018,Spanuth2020}, decentralize decision-making \citep{Robertson2015}, self-organize, be autonomous \citep{Büyükboyaci2019,Mollet2022,Puranam2014}, and establish fluid boundaries among them \citep{Bell2017}. These requirements often result in highly dynamic teams whose composition changes frequently, with their members periodically moving in and out \citep{Bell2017,Tannenbaum2012}. Many organizations have significantly altered their structure to adapt their internal operations to these demands, leading to the emergence of new organizational forms \citep{Burton2018}. For instance, virtual or distributed organizations employ communication technologies to connect their members even when they are geographically dispersed \citep{Squicciarini2011}. This reliance on communication technologies allows them to quickly alter their teams' structure in response to present and future demands \citep{Squicciarini2011}. Conversely, holacracies are decentralized organizations that favour autonomous team formation and shifts in composition based on the team members' self-assessment of their capabilities \citep{Robertson2015}. Similarly, platform-based organizations combine the use of communication technologies with self-organization, functioning as a series of autonomously-formed teams that change their composition in response to what the organization requires \citep{Burton2018}. Finally, organizations structured around project-based teams emphasize the creation of temporal groups of people that disband upon completing a specific task or one part of it, with its members moving on to other teams afterwards \citep{Lundin1995}. Consequently, we may argue that these new organizational forms often emphasize dynamic team composition as a conscious design element \citep{Burton2018,Tannenbaum2012}.

\cite{Burton2018} define organizational design as \textit{"a systemic approach to aligning structures, processes, leadership, culture, people, practices, and metrics to enable organizations to achieve their mission and strategy."} Each formal design element---or choice---is a configuration of these components. While prior research often overlooks the role of dynamic team composition as a design choice, there has been a growing interest in investigating how dynamic team composition can be a structural feature of organizations and how it may relate to other design choices and task performance \citep{Bell2017,Mathieu2014,Tannenbaum2012}. For instance, prior research suggests that organizations employ dynamic team composition to broaden the knowledge base of their teams and members \citep{Bell2017,March1991,Simon1991}, ensure that teams are well-adapted to the task requirements \citep{Bell2017,Mathieu2014,Spanuth2020,Tannenbaum2012}, become more innovative \citep{Bell2017,Spanuth2020}, and improve team-building efficiency over time \citep{Bell2017,Mathieu2014,Tannenbaum2012}. Moreover, there is evidence of the effects of dynamic team composition that contradicts the traditional perspective on the matter. For instance, \cite{Spanuth2020} argues that dynamic team composition may have a positive impact on task performance. These positive effects, however, may only manifest themselves if team members properly coordinate their actions \citep{Spanuth2020}.

This conflicting evidence between the rather traditional and modern branches of organizational research shows that shifts in the characterization of dynamic team composition may alter our understanding of its relationship with coordination and task performance. Given this conflicting evidence, our aim is to clarify how these factors---dynamic team composition, coordination, and task performance---interrelate. To achieve this objective, we characterize dynamic team composition as a design element consciously implemented by organizations and examine how it interacts with selected coordination modes, influencing task performance. From a practical standpoint, this paper aims at providing practical advice for teams and organizations regarding their coordination mode choices based on our results.

Design elements---such as dynamic team composition and coordination modes---have different effects depending on the organizational context, i.e., on a multidimensional set of components related to the organization such as its goals, strategies, structure, external environment, tasks faced, and human resources \citep{Burton2005,Burton2018}. This argument follows the \textit{contingency theory}, which suggests that there are no universally optimal organizational structures \citep{Galbraith1973}. Therefore, to properly characterize dynamic team composition as a design element and examine its interrelations with coordination, it is essential to consider the various contextual factors that may be related to them. %Instead, the effect of each individual design element needs to be examined in its particular context \citep{Burton2005,Burton2018,Galbraith1973,Mintzberg1979,Siggelkow2005}. 

The interrelations between dynamic team composition and coordination and their joint effects on task performance may be contingent on contextual factors that do not form part of the organization's design, such as individual learning \citep{Burton2018,Carley1992,Carley1996}. We understand individual learning as the process by which individuals discover new solutions and forget redundant solutions to a task over time \citep{Roth1995,Miller2016}. According to the traditional perspective on dynamic team composition, shifts in team composition trigger adaptation, forcing teams and their members to adapt their knowledge to these changing conditions \citep{Carley1992,Lin1992,Lin2006}. Conversely, dynamic team composition may be employed by teams to acquire previously unavailable knowledge \citep{Bell2017,March1991,Savelsbergh2015,Simon1991}. According to prior research, teams acquire new knowledge either through changes in their composition or the individual learning process of their members \citep{March1991,Simon1957}. Therefore, given standard categorizations of learning, dynamic team composition can be regarded as an adaptation process similar to learning at the team level. %While prior research has acknowledged the interrelations between dynamic team composition and individual learning, most of it studies the relationship between dynamic team composition and collective-level learning. Furthermore, prior findings are inconclusive, associating dynamic team composition to both negative \citep{Carley1992,Edmondson2001,Lewis2007,Reagans2005,Savelsbergh2015} and positive effects \citep{Mortensen2014,Sergeeva2018} on team-level learning. These mixed results show that 
Since both individual learning and dynamic team composition appear to be highly relevant for the adaptation process of the organization, investigating how these two processes may interact becomes crucial. Furthermore, prior research suggests interrelations between coordination and learning \citep{Edmondson2001,Reagans2005}. Specifically, researchers suggest that when team members coordinate properly, learning routines emerge, improving their knowledge acquisition process \citep{Edmondson2001,Reagans2005}.

Task complexity is also a contextual factor that may influence the effects of dynamic team composition and coordination on task performance \citep{Burton2018,Rivkin2003,Siggelkow2005,Wall2018}. Task complexity refers to the interdependencies among the several decisions which form an overall task \citep{Giannoccaro2019,Wall2016}. According to prior research, interdependencies increase uncertainty in decision-making and decrease task performance  \citep{Levinthal1997,Lin1992,Wall2016,Wall2018}. To reduce this uncertainty, the mirroring hypothesis argues that an organization's design should \textit{fit} these interdependencies \citep{Galbraith1973,Wall2018}. From a team's perspective, prior research suggests that team composition should be chosen according to the interdependencies among the decisions of the task \citep{Hsu2016}. Additionally, proper coordination among team members should be assured to further reduce this uncertainty \citep{Galbraith1973}. Thus, task complexity, coordination, and team composition appear to be interrelated.%Nevertheless, the lack of extensive research on the effects of dynamic team composition on task performance highlights a significant research gap in the existing literature.

In summary, our objective is to clarify the relationship between dynamic team composition, coordination, and task performance. Furthermore, and building on prior research \citep{Blanco-Fernandez2023,Blanco2023}, we consider two factors that might influence this relationship: individual learning \citep{March1991,Simon1991} and task complexity \citep{Siggelkow2005,Wall2018}. To achieve our objective, we have implemented an agent-based modeling approach. Agent-based modeling approaches are particularly helpful for studying organizational design, as they allow researchers to simultaneously model and study numerous design elements that interact with each other and other factors \citep{Burton2018,Leitner2021,Rivkin2003,Siggelkow2005,Wall2018}. Furthermore, simulations are especially valuable when data is unavailable or difficult to obtain \citep{Burton2018,Leitner2015,Wall2020}. Specifically, we follow previous research on organizational design by implementing a simulation model based on the NK framework \citep{Leitner2021,Rivkin2003,Siggelkow2005,Wall2018}. In our approach, a population of human decision-makers with limited capabilities---the \textit{agents} of the model---must form a team to solve tasks that we assume to differ in their level of complexity. This team may change its composition periodically. Additionally, the team members may employ a specific coordination mode when making their decisions to solve the complex task \citep{Siggelkow2005}. Furthermore, we assume that the agents of the model may learn about the complex task over time, adapting their knowledge to its conditions.

We believe the NK framework is appropriate for answering our research questions for several reasons. First, at its core, the NK framework involves simulating and observing over time the behavior of learning agents concerning specific interdependent elements \citep{Levinthal1997,Leitner2015,Wall2018}---in our case, the decisions of a complex task. Second, the NK framework can be employed to simulate and observe decision-making in small-scale teams over an extended period---see, for instance, \cite{Giannoccaro2018,Hsu2016,Rivkin2007,Siggelkow2005} and \cite{Wall2018}. Finally, the model is flexible enough to introduce variations in team composition \citep{Blanco-Fernandez2023,Blanco2023,Blanco-Fernandez2023a}. 

%This paper is structured as follows: Section \ref{sec:background} discusses the background on dynamic team composition coordination, and organizational design. We introduce the model in Sec. \ref{sec:model}. Results are provided and discussed in Sec. \ref{sec:results}. Finally, Sec. \ref{sec:conclusion} concludes the paper.

\section{Background}\label{sec:background}
\subsection{Dynamic team composition and organizational design}

Prior research on dynamic team composition may be divided into two diverging traditions that lead to seemingly contradictory insights and suggestions. In the branch of research concerned with more traditional organizations, teams are assumed to be stable units that aim at maintaining the same composition over time \citep{Tannenbaum2012}. This line of research usually associates dynamic team composition to organizational turnover, which is often interpreted as a cause of organizational crises and stress \citep{Lin1992,Lin2006}. This negative characterization of dynamic team composition is partly supported by prior findings. Simulation-based research suggests that the performance of teams experiencing turnover is significantly lower than that of  stable teams \citep{Carley1992,Lin1992}. These insights are also supported by experimental studies \citep{Harrison2003,Kim2014,Littlepage1997,Rao2006} and in survey-based research in the consulting \citep{Gardner2010}, healthcare \citep{Reagans2005}, and software development \citep{Mortensen2014} sectors. Additionally, simulation experiments in this branch of research suggest a negative association between dynamic team composition and organizational learning \citep{Carley1992}. Evidence coming from surveys in Dutch firms \citep{Savelsbergh2015} and the US healthcare sector \citep{Edmondson2001,Reagans2005} is in line with these insights.

The main mechanism that drives these negative effects is the increased uncertainty in decision-making brought by shifting team composition. According to \cite{Lin2006}, increased turnover may disrupt the organizations' operating conditions, intensifying uncertainty. This increased uncertainty forces decision-makers to act under suboptimal conditions \citep{Lin1992} and severely restricts the team's ability to preplan and anticipate future needs \citep{Galbraith1973}. This reduces the team members' decision-making capacity, decreasing task performance \citep{Galbraith1973,Lin2006}. %Hence, dynamic team composition carries a negative connotation in this line of research, with prior studies aiming at finding the causes of turnover and the mechanisms that may reduce it---see, for instance, \cite{Carley1992,Carley1998,Christian2017,Davis1973,Iammartino2016,Jungyoon2014,Lin1992} and \cite{Lin2006}.}

This branch of research, however, fails to acknowledge that dynamic team composition may be a conscious design element implemented by organizations rather than a consequence of suboptimal design choices. \cite{Tannenbaum2012} argue that the emergence of new organizational forms implies that research should depart from the traditional understanding of dynamic team composition. According to their arguments, many modern organizations---such as holacracies, virtual organizations, or project-based teams---rely on self-organized teams \citep{Mollet2022} with decentralized decision-making and shifting roles, functions, and personnel \citep{Robertson2015}. They also emphasize periodical changes in team composition as a crucial design element \citep{Tannenbaum2012,Webber2004}. This dynamic team composition allows organizations to learn how to configure the most effective possible team over time \citep{Bell2007,Mathieu2014} and enables teams to adapt to the task's conditions at any given time by changing their composition \citep{Savelsbergh2015}. This helps teams to integrate new members with previously unavailable knowledge within their ranks \citep{Simon1991}.

The research branch that follows these suggestions provides findings that challenge the traditional perspective on dynamic team composition and its supporting evidence. For instance, prior simulation-based results in small-scale teams suggest that task performance might increase when teams change their composition periodically \citep{Blanco-Fernandez2023,Blanco2023}. These insights are supported by experimental studies. For instance, \cite{Choi2004} and \cite{Choi2005} show that shifts in team composition enhance innovativeness in decision-making, increasing task performance. Survey-based research of German firms \citep{Spanuth2020} and Australian R\&D teams \citep{Hirst2009} confirm these findings. This increase in innovativeness may result from faster learning \citep{Sergeeva2018} and from better transfer of knowledge between team members \citep{Choi2005,Sergeeva2018}. Thus, dynamic team composition can be understood as an exploratory process by which teams acquire new knowledge \citep{March1991,Mortensen2014,Simon1991}. Exploratory processes are particularly valuable for enhancing the performance of complex tasks, particularly during the initial stages of decision-making \citep{Levinthal1997}.%Therefore, interpreting dynamic team composition as a concious design choice leads to different insights to those brought by following the traditional perspective on the matter

In summary, both branches of research have brought forward the main advantages and disadvantages of dynamic team composition. On the one hand, dynamic team composition may be associated with enhanced innovativeness \citep{Choi2005} and knowledge transfer \citep{Sergeeva2018}, which may lead to improvements in performance \citep{Blanco-Fernandez2023,Blanco2023}. On the other hand, periodical shifts in team composition may increase uncertainty in decision-making, reducing task performance \citep{Lin2006}. Modern organizations that rely on dynamic team composition might seek to capitalize on these advantages while adopting design elements that aim at reducing its associated uncertainty. For instance, prior research argues that implementing effective coordination mechanisms among team members may successfully counteract uncertainty in decision-making  \citep{Galbraith1973}.

\subsection{Organizational design and coordination modes}

\cite{Galbraith1973} suggests that organizations should alter their design when facing significant uncertainty, either by enhancing their information processing capacity or by decreasing the demand for it. Organizational design elements are often divided into structure and coordination choices \citep{Burton2018,Leitner2021,Leitner2023}. Structure choices are those related to the division of labor \citep{Burton2018}, and they are typically used to reduce the information processing demands of the organization \citep{Galbraith1973}. Generally, organizations perform better if their structural features eliminate the interdependencies among different decision-makers or reduce them to their greatest possible extent \citep{Wall2018}. This objective is relatively easy to achieve for simple tasks with few concentrated interdependencies. If tasks are complex---i.e., interdependencies are numerous and widespread throughout the task---achieving this objective becomes much more difficult \citep{Burton2018,Siggelkow2005,Wall2018}. %This increases uncertainty for decision-makers which, in turn, may have negative consequences for task performance \citep{Wall2018}. Consequently, organizations cannot rely exclusively on structure choices to improve task performance in complex settings \citep{Burton2018}.

Hence, organizations might resort to coordination choices to overcome the limitations caused by task complexity \citep{Burton2018}. Coordination choices encompass the actions taken to facilitate the flow and exchange of information between individuals, thereby increasing the  organization's information processing capacity \citep{Burton2005,Galbraith1973,Leitner2020}. According to \cite{Nadler1997}, coordination choices include informal coordination through the establishment of \textit{incentive schemes} and formal coordination through the implementation of \textit{coordination modes}\footnote{Coordination choices and coordination modes are two separate concepts. Coordination choices are part of organizational design and refer to those elements which ensure that organizations combine their resources efficiently. The choice of a proper coordination mode is a design problem in itself and part of the coordination choices of the organization \citep{Burton2018}.}, i.e., through the implementation of rules and procedures which organize individual decision-making and assure open communication between decision-makers \citep{Burton2018,Leitner2022,Siggelkow2005,Wall2018}.

%According to prior research \citep{Leitner2021,Rivkin2003,Siggelkow2005,Wall2017}, team members may align their actions with the organization's objectives} by implementing a specific incentive scheme. Specifically, organizations aim to align their own goals with those of their members} by implementing a specific incentive scheme that rewards decision-makers based on the overall task performance rather than individual achievements \citep{Siggelkow2005}. The underlying assumption behind this mechanism is that by aligning objectives, the individual decision-makers will align their actions, too \citep{Leitner2022,Rivkin2003}. Prior research suggests ambiguous effects of these informal coordination mechanisms}. While there are results that suggest positive effects of task-oriented incentives on complex task performance, other findings indicate that these incentives reduce task performance in the long run \citep{Leitner2021,Rivkin2003,Siggelkow2005}. 

Prior research has highlighted the importance of implementing a particular coordination mode---i.e., of establishing formal links, procedures and communication channels---to formally coordinate the team members' actions \citep{Galbraith1973,Mintzberg1979,Nadler1997}. Each coordination mode has its specific functioning and characteristics. For instance, \cite{Nadler1997} argue that when decisions are taken sequentially, decision-makers require substantial information about the preceding actions to act properly. Previous simulation-based research tested this hypothesis by studying a coordination mode labeled as \textit{sequential decision-making}, suggesting that it may improve performance if tasks are sufficiently complex \citep{Wall2017,Wall2018}.

Conversely, establishing communication channels among team members is also emphasized as an effective tool for achieving coordination \citep{Galbraith1973,Siggelkow2005}. Specifically, \cite{Mintzberg1979} highlight two different ways of achieving coordination following this approach. First, coordination may be achieved by mutual adjustment, i.e., by establishing direct, unmoderated lateral communication between team members. According to \cite{Galbraith1973}, this direct interaction ensures that team members make their decisions together. \cite{Siggelkow2005} label this coordination mode \textit{lateral communication}. Their results suggest that, for complex tasks, lateral communication counteracts the uncertainty associated with task complexity better than any other coordination mode, improving task performance \citep{Siggelkow2005}. Second, coordination may be achieved by assuring direct, moderated communication between team members through the establishment of liaison devices \citep{Galbraith1973,Mintzberg1979,Nadler1997}. Liaison devices encompass various roles which, while having no decision-making capacity of their own, are tasked with assuring that information flows properly between team members \citep{Mintzberg1979,Nadler1997}. Team members act autonomously and fully retain their decision-making capacity, while the liaison device ensures that they maintain communication \citep{Nadler1997,Siggelkow2005}. Although they may effectively achieve their objective, prior simulation-based results suggest that a \textit{liaison coordination mode} may slow down decision-making, reducing task performance in the long run \citep{Siggelkow2005}. 

\subsection{Dynamic team composition and coordination}
Prior research often addresses the interrelations between dynamic team composition and coordination \citep{Edmondson2001,Edmondsonetal2003,Gardner2010,Harrison2003,Mortensen2014,Reagans2005}. There are, however, diverging interpretations of the interactions between dynamic team composition and coordination. Evidence coming from one branch of research suggests that dynamic team composition is negatively associated with coordination. For instance, experimental findings suggest that shifts in composition reduce coordination among team members and, in turn, task performance \citep{Harrison2003}. Survey-based findings in the fields of healthcare \citep{Edmondson2001,Edmondsonetal2003,Reagans2005}, software development \citep{Mortensen2014}, and consulting \citep{Gardner2010} support these experimental results. These findings, however, may be constrained by the traditional perspective on dynamic team composition, which associates it to organizational turnover \citep{Mathieu2014}.

As dynamic team composition gains significance as an organizational design element, it is likely that new insights will be found if research departs from the traditional perspective on dynamic team composition, interpreting it rather as part of organizational design \citep{Tannenbaum2012}. For instance, the investigation of German firms carried out by \cite{Spanuth2020} leads to an alternative interpretation on the relationship between dynamic team composition and coordination. Specifically, their results suggest that dynamic team composition enhances creativity and innovation within teams and improves task performance. Nevertheless, proper coordination among team members needs to be ensured if teams want to fully exploit these advantages \citep{Spanuth2020}.

\section{The Model}\label{sec:model}
In this paper, we build on the NK framework \citep{Levinthal1997} and propose an agent-based model to study the effect of coordination modes and dynamic team composition on task performance. In contrast to experimental and empirical research, our proposed model does not rely on observing real-life data \citep{Wall2020}. Instead, we aim to better understand the variables of interest by employing a simulation approach that develops the insights found in prior research (see Sec. \ref{sec:background}). Three factors influence our choice of the research method. First, the NK framework provides an opportunity for creating an artificial organization that we can use to change its design choices and examine their effects on task performance \citep{Wall2016,Wall2020}. Additionally, our adaptation of the NK framework helps us explore the long-term dynamics of team composition \citep{Wall2020}, which is in line with the suggestions given in prior research \citep{Mathieu2014}. Second, we can modify the NK framework to model aspects such as learning, task complexity, and agent interaction \citep{Wall2020}. Finally, the NK framework is a standardized approach employed in related research \citep{Baumann2019,Wall2016} and contrasted with other methods \citep{Billinger2014,Giannoccaro2018}. Consequently, we believe that this model is the best choice for our research.

\begin{figure}[ht]
    \centering
    \includegraphics[width=1\textwidth]{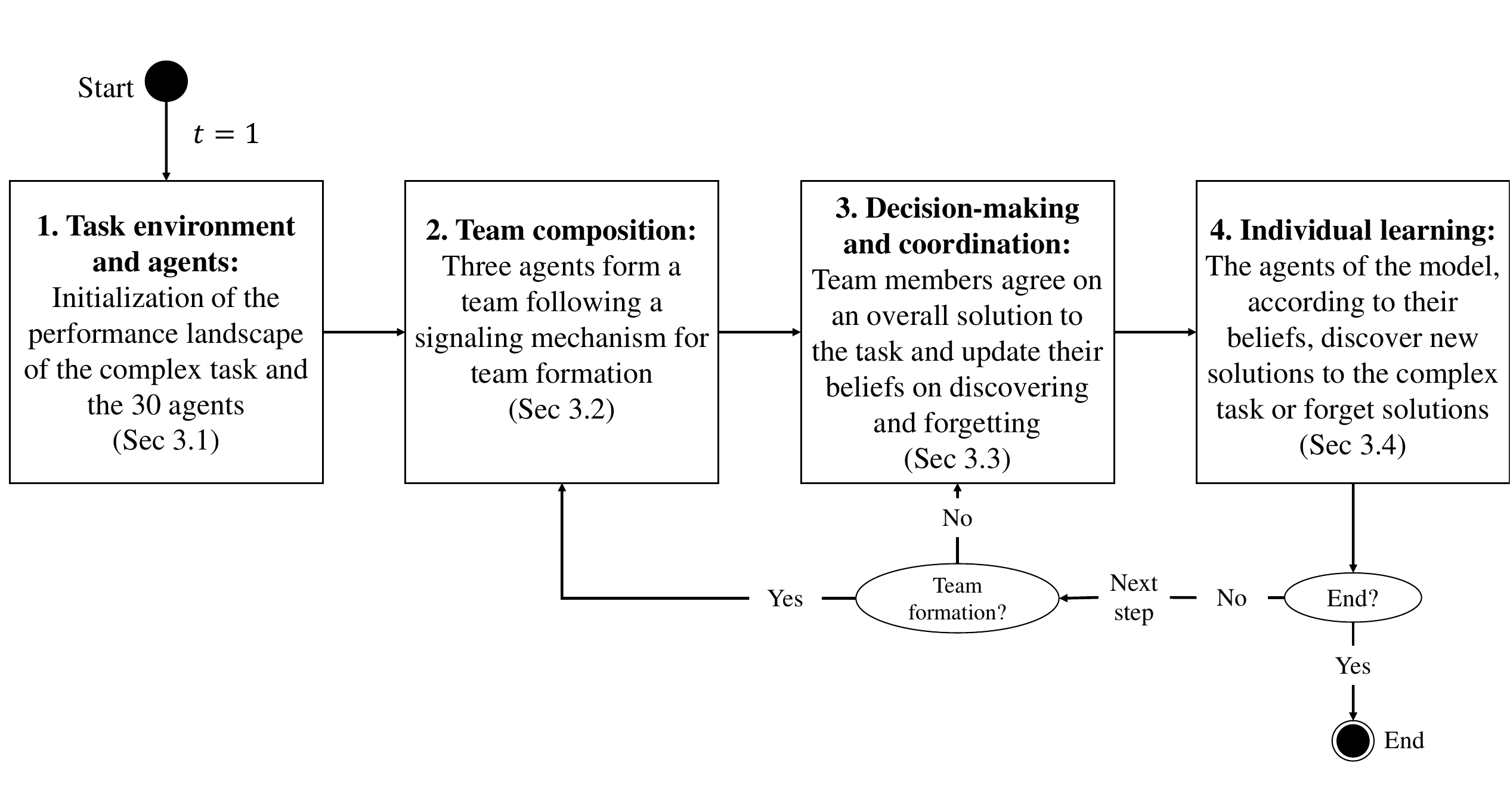}
    \caption{Sequence of the model.}
    \label{fig:sequence}
\end{figure}

The following four subsections  correspond to the four building blocks of the model and follow the sequence of events during each simulation (see Fig. \ref{fig:sequence}). In this agent-based model, several human decision-makers are entrusted with solving a complex task (see Sec. \ref{sec:environment}). These agents must form a team to solve this complex task; the process of team formation is repeated periodically to account for dynamic team composition (see Sec. \ref{sec:team}). Once the team is formed, team members follow a specific coordination mode to implement a solution to the complex task (see Sec. \ref{sec:decision}). At the end of each period, all agents update their knowledge by going through a learning process (see Sec. \ref{sec:learning}). 
Subsequently, depending on the parameter configuration, the process returns to either team formation or the decision-making phase.

Section \ref{sec:variables} discusses the parameter settings. Appendix \ref{app:technical} provides a discussion on the technical aspects of the model and the simulation.

\subsection{Task environment and agents}\label{sec:environment}
We model the complex task as a vector of $N=12$ binary decisions. Each decision, denoted by $d_{n}$, contributes $c_{n}$ to the overall task performance, which we calculate as the average of all contributions (see Eq. \ref{eq:taskperf} and Appendix \ref{app:landscape}). Each vector of $12$ binary values is a \textit{solution} to the complex task and has an associated task performance. The mapping of each of the $2^{12}=4,096$ possible solutions to their associated performance is the \textit{performance landscape} \citep{Levinthal1997} (see Appendix \ref{app:landscape} for additional details). Following the NK framework, we incorporate \textit{task complexity} by making the decisions \textit{interdependent} \citep{Giannoccaro2019,Levinthal1997,Wall2016}. Thus, the contribution to task performance $c_{n}$ depends on the decision itself $d_{n}$ and $K$ other decisions. A larger value of $K$ indicates more interdependencies and, thus, a higher task complexity. When there are no interdependencies, i.e., $K=0$, only one optimal solution to the complex task exists, which is easily achievable. As $K$ increases and approaches $N-1=11$, several suboptimal solutions appear, making it increasingly challenging to reach the optimal solution \citep{Levinthal1997,Rivkin2007}.

Furthermore, task complexity has significant implications for the decision-making agents. To solve a complex task, it is often necessary to combine different complementary skills that are typically not found in a single individual \citep{Hsu2016}. Consequently, a single agent cannot solve a complex task by themselves, as they have \textit{limited capabilities} concerning the complex task they face \citep{Funke1995}. Instead, complex tasks are usually solved by teams of different experts \citep{Giannoccaro2019,Hsu2016,Rivkin2003}. To account for these limited capabilities, we divide the $12$-dimensional task into $M=3$ subtasks, each consisting of four decisions. Each subtask corresponds to an area of expertise \citep{Hsu2016}. We randomly assign each agent of the model to one of these areas of expertise, i.e., they can only solve one of the three subtasks. Additionally, agents are not aware of the complete set of solutions to their assigned subtask from the beginning. Instead, we initially endow each agent with one random solution to their assigned subtask. This characterization of limited capabilities follows the \textit{difficulty assumption} as described by \cite{Licalzi2012}, ensuring that agents do not have the complete set of solutions to their assigned subtask at their disposal from the beginning.

There are $P=30$ agents in our model. Agents are utility-maximizing, and they experience utility by participating in the team and solving the complex task. A team member's utility is the average of their own performance contributions and the remaining members' performance contributions---i.e., the performance contributions of the \textit{residual decisions.} (see Eq. \ref{eq:utility} and Appendix \ref{app:agents}) In addition to the aforementioned limited capabilities, agents also have restricted cognitive capacities, which prevent them from pursuing long-term objectives. Instead, agents are myopic and aim at maximizing their immediate utility.\footnote{This ensures that agents behave analogously to a hill-climbing process on the performance landscape. This is a defining feature of the NK framework, see Appendix \ref{app:technical} for additional details.}

\subsection{Team composition}\label{sec:team}
We assume that one team of three members---one per area of expertise---is sufficient to solve the entire complex task.\footnote{Team member $1$ deals with decisions $1$ to $4$, member $2$ with decisions $5$ to $8$, and team member $3$ with decisions $9$ to $12$.} Following the increasing reliance on self-organization that characterizes modern organizations \citep{Puranam2014,Robertson2015}, we set up a team formation mechanism in which agents autonomously come together to create the team. As \cite{Robertson2015} notes in his study of holacracies, members of modern organizations often decide autonomously whether to join and leave teams by assessing their own expertise and potential contributions. Additionally, team members often take a proactive stance towards self-improvement and the upgrade of the team's operations \citep{Ngo2023}. To apply this modern approach to self-organization, the agents in our model follow a signaling mechanism for team formation that aims at integrating the best available experts within the team \citep{Blanco-Fernandez2023,Blanco2023}. 

We work on certain assumptions about the agents' behavior to avoid any strategic behavior. First, we assume that agents know perfectly how the team composition is chosen. Additionally, we assume that agents do not cheat during team formation. Finally, we omit any communication between agents during the team formation process. 

Team formation works as follows: Each agent, who knows a specific set of solutions, estimates the utility associated with these solutions (see Eq. \ref{eq:estutility} and Appendix \ref{app:team}). Agents cannot, however, anticipate the residual decisions for the incoming period or communicate to obtain information on them, since these decisions are located outside their area of expertise. Instead, they assume that the residual decisions from the previous period will not change for the incoming period. Furthermore, agents may commit small miscalculations which range between 1 and 10 percent, reflecting the average error rate in reporting according to prior research \citep{Tee2007}. Each agent then sends a signal equivalent to the highest estimated utility available. The agent with the highest signal for each subtask joins the team.\footnote{In the case of a draw, the team member is randomly chosen from the top signalers.} Consequently, the team is formed by the three agents who are supposed to be the best-available experts in their respective subtasks.\footnote{Non-members wait on the sidelines until the next team formation process, earning utility equal to 0.}

The team is always formed in the first period. Afterwards, agents periodically change the team composition by repeating the team formation process every $\tau$ periods. $\tau$ is a design parameter that can be understood in terms of the team's lifetime: A higher (lower) value of $\tau$ indicates that team formation occurs less (more) frequently, which implies a long-term (short-term) team composition. %Consequently, we can interpret $\tau$ as a design parameter which organizations may use to control the team composition.

\subsection{Decision-making and coordination modes}\label{sec:decision}
Once the team is formed, the three team members propose solutions to their assigned subtasks. These proposals are combined to create an overall joint solution to the complex task. %There are, however, many ways by which decision-makers within teams may arrive at a joint solution. 
Following prior research \citep{Siggelkow2005,Wall2018}, we define three coordination modes. The choice of the coordination mode determines how team members evaluate their solutions, which criteria they follow to make their proposals, how these proposals are shared between team members, how proposals are combined to form candidate solutions, and how team members agree on implementing an overall task solution \citep{Burton2018,Siggelkow2005,Wall2018}. These coordination modes differ in three aspects \citep{Siggelkow2005}: The team members might

\begin{itemize}
    \item evaluate their solutions privately before making proposals.
    \item control in which order their proposals are considered.
    \item have veto power over the remaining members' proposals.
\end{itemize}

\noindent Additionally, we define a benchmark scenario without coordination. In this benchmark scenario, we study a \textit{fully autonomous} team in which decision-making is entirely decentralized, without communication between team members \citep{Siggelkow2005,Wall2018}. This highly decentralized structure corresponds to a congregation of loosely connected individuals with mutual interests \citep{Horling2004} that can be found, for example, in professional bureaucracies like schools \citep{Mintzberg1979}. Furthermore, this fully autonomous team reflects the need for rapid action and innovation that drives many organizations. It also reflects their strong reliance on full self-organization and decentralization \citep{Nadler1997}. In this scenario, each team member estimates the utility of each solution they know as outlined in Sec. \ref{sec:team} (see Eq. \ref{eq:estutility} and Appendix \ref{app:team}). Then each member chooses the solution to their subtask which reports the highest estimated utility, and the overall task solution for the current period is the concatenation of all members' proposals.

Teams often attempt at simplifying the process of solving complex tasks by partitioning them into multiple steps and addressing them sequentially \citep{Burton2005,Mintzberg1979}. When there are corresponding interdependencies among team members, this creates a sequential reliance among them, as they must be aware of prior decisions to act accordingly \citep{Mintzberg1979,Nadler1997}. This is evident, for instance, in ass production factories \citep{Mintzberg1979} or in activities such as the processing of checks in banks \citep{Nadler1997}. Following \cite{Wall2018}, we transfer these insights onto the \textit{sequential} coordination mode. In this coordination mode, the proposals are made successively according to the subtask order \citep{Wall2018}. This coordination mode, thus, includes individual evaluation, but members cannot veto other proposals or control the order of choice. The team member associated with the first subtask makes a proposal and reports it to the following members, who update their residual decisions accordingly. This process is repeated for each member until they all have made their proposals. Thus, the team member associated with the first subtask chooses according to previous period residual decisions, while the remaining team members choose according to their updated residual decisions. The overall solution is then the concatenation of all these sequential proposals.

Organizations that rely on decentralized decision-making and self-organization often encounter challenges when establishing communication channels among their members \citep{Mintzberg1979}. Consequently, liaison devices are particularly relevant in modern organizations such as holacracies \citep{Robertson2015} and adhocracies \citep{Siggelkow2005}. A liaison device involves establishing a mediator which handles communication among team members, who in turn retain their decision-making power \citep{Galbraith1973}. Liaison devices rely on direct meetings between team members to discuss their current information and future intentions \citep{Mintzberg1979}. We follow \cite{Siggelkow2005} and model these insights in the \textit{liaison} coordination mode, which retains decentralized decision-making while introducing direct communication between the team members. The liaison coordination mode includes individual evaluation, veto power, and control of the order in which proposals are considered, and works as follows \citep{Siggelkow2005}. Each member ranks each solution known in terms of their estimated utility. Then they present their two highest-ranked proposals in a coordination session.\footnote{As \cite{Siggelkow2005} show, the number of proposals selected by decision-makers does not improve performance significantly for a liaison coordination mode.} Two candidate solutions are formed, one by concatenating the team members' preferred proposals and other by concatenating their second-preferred proposals.\footnote{\cite{Siggelkow2005} show that increasing the number of candidate solutions that decision-makers evaluate improves performance significantly for a liaison coordination mode. We keep the candidate solutions restricted to two to retain the same processing power for the liaison and lateral communication coordination modes, see footnote 8 later on.} Each team member evaluates the two candidate solutions according to their estimated utility. Team members accept the candidate solution if its estimated utility is higher than the last achieved utility. Otherwise, they veto the candidate solution. If both candidate solutions are vetoed, the overall solution from the previous period remains unchanged. Conversely, if all team members accept one candidate solution, it becomes the overall solution for the upcoming period.

Since fast decision-making and innovation is often demanded in organizations, teams may reject the establishment of liaison devices while retaining broad communication among their members \citep{Nadler1997}. Thus, team members may coordinate their actions by mutual adjustment, i.e., by communicating directly and without moderation \citep{Mintzberg1979}. Mutual adjustment implies that team members retain their decision-making capacity, but they must exchange information before taking any action \citep{Mintzberg1979}. According to \cite{Galbraith1973} this results in team members jointly agreeing on decisions. These insights are transferred onto the \textit{lateral communication} coordination mode \citep{Siggelkow2005}. This coordination mode retains the team members' veto power, but they cannot evaluate their solutions privately or consider the proposals in their preferred order \citep{Siggelkow2005}. The lateral communication mode works as follows: Each member randomly chooses two solutions to their assigned subtask from the set of known solutions.\footnote{According to \cite{Siggelkow2005}, the number of proposals selected by team members does not impact performance significantly for a lateral communication mode.} The team members take their two proposals to a coordination session. In this session, two candidate solutions are formed by randomly concatenating the team members' proposals.\footnote{In contrast to the liaison mode, the number of candidate solutions evaluated by team members does not impact performance significantly for a lateral communication mode \citep{Siggelkow2005}. We keep the number of candidate solutions evaluated to two to retain the same processing power for the lateral communication and liaison coordination modes, see footnote 6 earlier on.} Similarly to the liaison mode, team members then veto or accept the candidate solutions depending on their estimated utility. The team implements a candidate solution if all members accept it. Conversely, if the members veto both candidate solutions, the overall solution remains unchanged from the previous period.

\subsection{Individual learning}\label{sec:learning}
Recall that, due to their limited capabilities, agents do not know the complete set of solutions to their assigned subtask from the beginning (see Sec. \ref{sec:environment}). Instead, agents gradually \textit{learn} about the complex task and adapt their knowledge---i.e., their set of available solutions---over time \citep{Funke1995}.\footnote{Learning occurs for all agents, team members or others.} According to \cite{Miller2016}, learning is a process limited by the amount of knowledge that individuals can retain. Thus, as agents discover new things, they tend to forget prior knowledge which was not considered relevant in that particular time frame \citep{Miller2016}. As \cite{Roth1995} argue, forgetting is a process in which the agents' knowledge gradually deteriorates. Hence, prior research suggests that learning should be considered the combination of multiple interrelated processes, including discovering, forgetting, and recalling, rather than just discovering \citep{Miller2016}. Our characterization of individual learning follows the insights given by \cite{Roth1995} and \cite{Miller2016} and consists of two processes: discovering and forgetting \citep{Blanco-Fernandez2023,Blanco2023}.\footnote{Agents may forget and rediscover the same solution multiple times during a simulation.}

Each agent may \textit{discover} a solution to their assigned subtask at the end of every period with probability $\mathbb{P}$. This solution is chosen randomly from all unknown solutions. Each solution has a certain likelihood of being discovered that is influenced by the agent's beliefs about it (see Appendix \ref{app:learning} for additional technical details). These beliefs are periodically updated according to their experience and following a Bayesian updating rule \citep{Leitner2021,Tosic2010}.

The updating process works as follows. Each time a team member implements a solution to their assigned subtask, they experience a particular utility. Then, the team member compares this utility with the utility experienced in the previous period. If this solution represents an improvement in their utility compared to the previous period, its likelihood of being discovered will increase. Conversely, this likelihood will decrease if it reduces the team member's utility compared to the previous period. We formalize this process in Eqs. \ref{eq:learn_a} and \ref{eq:learn_b} in Appendix \ref{app:learning}.\footnote{Non-member agents cannot update their solutions regarding the discovery process, as they cannot implement them.}

Thus, a belief is only updated when an agent  knows and implements a solution to their assigned subtask. Beliefs, however, specifically affect the likelihood with which solutions are discovered and are only relevant to unknown solutions. %Hence, agents should be able to \textit{forget} solutions over time, or these beliefs would be irrelevant.

Agents may forget one known solution at the end of each period with the same value of probability $\mathbb{P}$. The forgotten solution is randomly selected from the set of known solutions. Again, agents form beliefs regarding these solutions that determine their likelihood of forgetting. This likelihood may change as time passes and agents update their beliefs. Each time a team member implements a solution to their assigned subtask, they update their associated belief by comparing the utility experienced in the current and the previous period. The likelihood of forgetting this solution will increase if it reduces their utility compared to the previous period. Conversely, if this solution improves the team member's utility, the likelihood of forgetting it decreases.

Additionally, forgetting is related to the passing of time \citep{Brenner2006}. We introduce a memory factor in the agents' updating process, which reflects that agents tend to forget solutions %not only because they have implemented them, but also 
because they have stored them in their memory for too long \citep{Roth1995}.  In our model, all agents---team members or not---update the beliefs of the solutions they know at the end of every period, increasing the likelihood of forgetting each solution as time passes. We provide the general updating rule for all beliefs regarding forgetting in Eqs. \ref{eq:forget_a} and \ref{eq:forget_b} in Appendix \ref{app:learning}. %This means that $\forall \hat{\mathbf{d}}_{mi} \in \mathbf{S}_{mt}: \lambda^{(i)}_{mt} = \lambda^{(i)}_{m\{t-1\}}+1$. 

In conclusion, by implementing solutions and updating their beliefs, agents ensure that they retain those solutions perceived as beneficial and discard those perceived as harmful or irrelevant.

\subsection{Parameters and scenarios}\label{sec:variables}
\begin{figure}[ht]
 \centering
 \includegraphics[width=\textwidth]{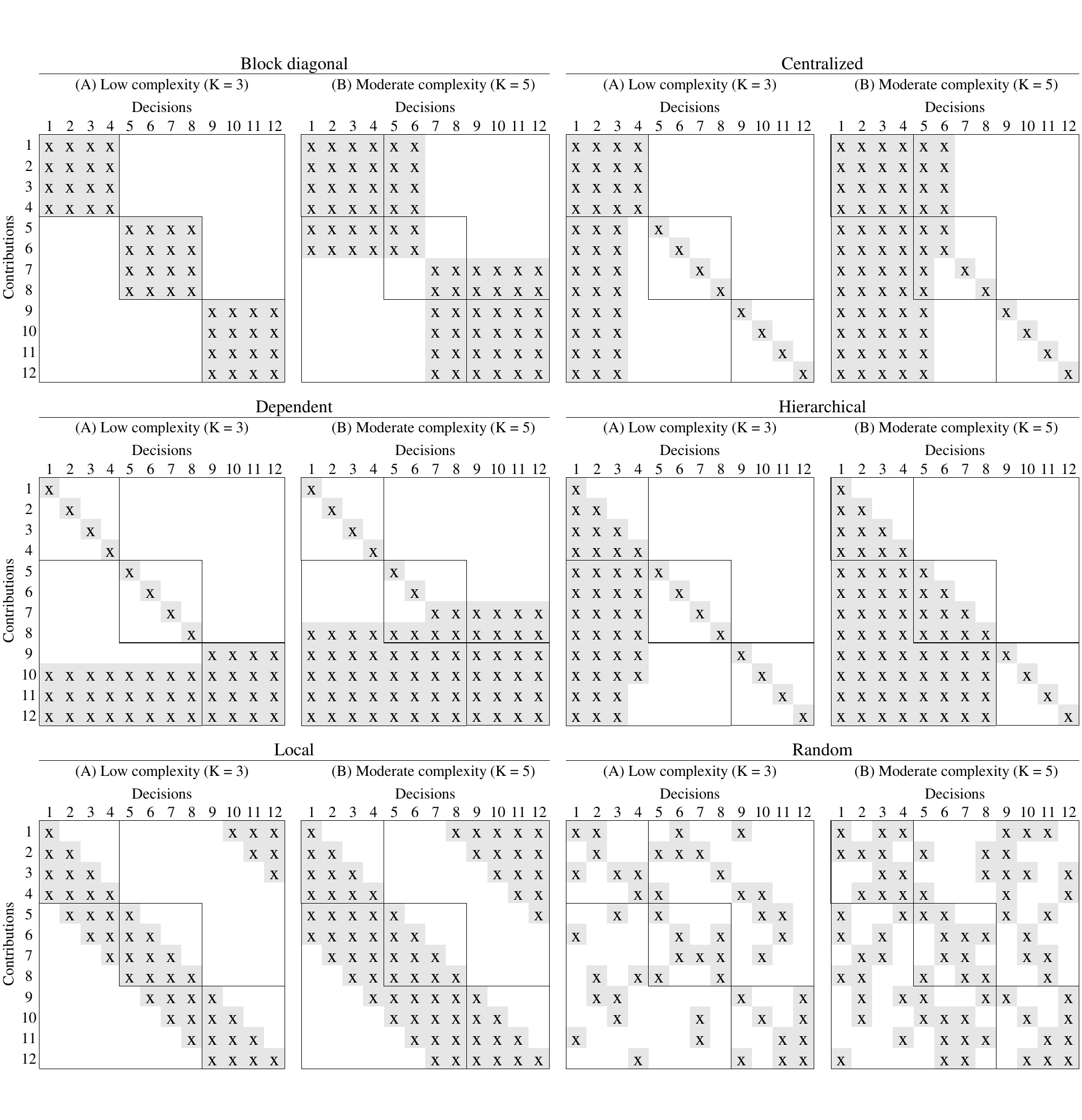}
 \caption{Interdependence matrices. Interdependencies between each contribution and each decision are highlighted in grey with an X.}
 \label{fig:matrix}
\end{figure}

As outlined in Sec. \ref{sec:environment}, we study a task formed by $N=12$ binary decisions divided into $M=3$ subtasks of length $S=N/M=4$. The group is formed by three team members out of a population of $P=30$ agents. Regarding task complexity, we consider two levels: Low ($K=3$) and moderate complexity ($K=5$). Additionally, we take into account six different \textit{interdependence structures}. These structures reflect which contributions depend on which decisions and, thus, the degree of task decomposability \citep{Rivkin2007} (see Appendix \ref{app:landscape} for additional details). According to prior research, the interdependence structure also determines the performance landscape's shape and, consequently, the team's performance \citep{Rivkin2007}. %Each structure has $N\times(K+1)$ interdependencies in total. 
We represent the six interdependence structures studied in Fig. \ref{fig:matrix}:

\begin{itemize}
    \item \textit{Block diagonal}: We allocate the interdependencies in squares of size $K+1$. For low complexity, this structure reflects a \textit{perfectly decomposable} task in which there are no cross-interdependencies between subtasks.
    \item \textit{Centralized}: Interdependencies are located within the first $K+1$ decisions. In low-complexity tasks, only the first team member influences the other members' contributions. In tasks of medium complexity, the power to influence others is shared between the first and the second member.
    \item \textit{Dependent}: This structure reverses the idea of the centralized structure: instead of making one team member very influential, we make one member highly dependent on the other members' decisions.
    \item \textit{Hierarchical}: Each decision influences the $N-n$ following contributions but not the preceding $n-1$.
    \item \textit{Local}: Each decision affects the following $K$ contributions. If there are no following contributions, interdependencies fall back to the beginning of the task (see Fig. \ref{fig:matrix}, Local).
    \item \textit{Random}: We randomly allocate the interdependencies throughout the task. Consequently, each member partly affects the other members' contributions.
\end{itemize}

\noindent Regarding dynamic team composition, we define $\tau$ in Sec. \ref{sec:team} as the number of periods between each instance of team formation. There are three scenarios depending on $\tau$:

\begin{itemize}
    \item A team with a \textit{long-term composition} maintains the same composition during the observation period. We represent this scenario by $\tau=\infty$.
    \item A team with a \textit{medium-term composition} changes its composition every $\tau=10$ periods.
    \item A team with a \textit{short-term composition} changes its composition in every period, so $\tau=1$.
\end{itemize}

\noindent Concerning individual learning, we set equal probabilities for discovering and forgetting. This reflects the agents' limited capabilities  and ensures that they discover and forget solutions at the same rate. This reduces the likelihood that each agent simultaneously knows all available solutions to their assigned subtask.\footnote{This is based on the difficulty assumption. More details are provided in \cite{Licalzi2012}.} In our analysis, we study probabilities between $\mathbb{P}=0$ and $\mathbb{P}=1$ in intervals of $0.1$. %This learning probability differs from the probability distribution of discovering and forgetting solutions. Specifically, the learning probability is common to all agents. In contrast, the probability distribution of discovering and forgetting solutions are unique for each agent and change over time depending on agent $m$'s beliefs.

We summarize the main parameters of the model and their values in Tab. \ref{tab:variables}. For our analysis, we simulate 1,584 scenarios, each consisting of 1,500 simulation rounds of 100 periods.\footnote{We fix the number of simulations following an analysis of the variance of the results; see \cite{Lorscheid2012}.}. The dependent variable in our analysis is the observed task performance at each period. To assure comparability between scenarios, we normalize this observed task performance by the maximum achievable performance in each simulation round. We then calculate the mean normalized performance at each period for the 1,500 simulation rounds. This measure is formalized in Eq. \ref{eq:mean-perf} in Appendix \ref{app:perf}.

\begin{table}[ht]
\caption{Parameters}
\label{tab:variables}
\renewcommand{\arraystretch}{1.2}
%\resizebox{\textwidth}{!}{%
\begin{tabular}{llll}
\\ \hline
Type                                  & Variables              & Notation                    & Values                         \\ \hline
\multirow{6}{*}{Independent variables} & Task complexity        & $K$                         & \{3, 5\}                       \\
                                      & Interdependence structure & \textit{Matrix} & See Fig. \ref{fig:matrix}     \\
                                      & Team composition      & $\tau$                      & \{$\infty, 1, 10$\}         \\
                                      & Learning probability   & $\mathbb{P}$                & $\{0:0.1:1\}$ \\
                                      & Time period             & $t$                         & $\{1:1:100\}$   \\
                                      & Coordination mode       & -                         & See Sec. \ref{sec:decision} \\\hline
Dependent variable                    & Task performance       & $C(\mathbf{d_t})$           & $[0,1]$                     \\ \hline
\multirow{5}{*}{Other parameters}       
                                      & Number of decisions    & $N$                         & 12                             \\
                                      & Population of agents   & $P$                         & 30                             \\
                                      & Number of subtasks     & $M$                         & 3                              \\
                                      & Simulation round  & $\phi$                         & $\{1:1:1,500\}$                       \\
                                      & Error term             & $e$ & $e\sim N(0,0.1)$ \\
\hline
\end{tabular}%
%} 
\end{table}

\section{Results}\label{sec:results}
This paper aims at examining the interrelations between selected coordination modes and dynamic team composition. Additionally, we investigate whether there are moderating effects of task complexity and individual learning. We present the results in the following three subsections. In Sec. \ref{subsec:general} we outline the general effects of dynamic team composition on task performance and its interrelations with the coordination mode implemented. In Sec. \ref{subsec:complex} we differentiate our results by task complexity and the interdependence structure and, finally, by individual learning in Sec. \ref{subsec:learning}.

\subsection{General effects of dynamic team composition}\label{subsec:general}
In this subsection, we report two performance measures: The mean performance of the $100$ periods of each simulation---see Eq. \ref{eq:meanperf} and Appendix \ref{app:perf}---and the final performance, denoting the performance achieved on average at $t=100$. Table \ref{table:generaleffects} reports the mean and final performances for each team composition and coordination mode studied. We also report the confidence limits below each performance at a 0.01\% confidence level and highlight in bold the highest performance achieved for each team composition considered.

\begin{table}[ht]
\caption{General effects of dynamic team composition}
\label{table:generaleffects}
\renewcommand{\arraystretch}{1.5}
\resizebox{\textwidth}{!}{%
\begin{tabular}{lccc|ccc}
                            & \multicolumn{3}{c|}{Mean performance}               & \multicolumn{3}{c}{Final performance}          \\ \hline
                            & Long            & Medium          & Short           & Long            & Medium          & Short           \\ \hline
\multirow{2}{*}{Fully autonomous}  & 0.8296          & 0.8531          & 0.8792          & 0.8435          & 0.8680          & 0.8921          \\
                            & \textit{($\pm$0.0029)} & \textit{($\pm$0.0030)} & \textit{($\pm$0.0026)} & \textit{($\pm$0.0029)} & \textit{($\pm$0.0030)} & \textit{($\pm$0.0026)} \\ \hline
\multirow{2}{*}{Sequential} & 0.8390          & 0.8668          & 0.8982          & 0.8510          & 0.8794          & 0.9095          \\
                            & \textit{($\pm$0.0022)} & \textit{($\pm$0.0026)} & \textit{($\pm$0.0022)} & \textit{($\pm$0.0022)} & \textit{($\pm$0.0026)} & \textit{($\pm$0.0022)} \\ \hline
\multirow{2}{*}{Liaison}    & 0.7601          & 0.7835          & 0.8120           & 0.7689          & 0.7987          & 0.8283          \\
                            & \textit{($\pm$0.0024)} & \textit{($\pm$0.0028)} & \textit{($\pm$0.0032)} & \textit{($\pm$0.0024)} & \textit{($\pm$0.0028)} & \textit{($\pm$0.0032)} \\ \hline
\multirow{2}{*}{Lateral}    & \textbf{0.8504} & \textbf{0.8774} & \textbf{0.9075} & \textbf{0.8816} & \textbf{0.9189} & \textbf{0.9410} \\
                            & \textit{($\pm$0.0019)} & \textit{($\pm$0.0020)} & \textit{($\pm$0.0018)} & \textit{($\pm$0.0019)} & \textit{($\pm$0.0020)} & \textit{($\pm$0.0018)} \\ \hline
\end{tabular}}
\end{table}

The results of Tab. \ref{table:generaleffects} suggest that more frequent team formation is generally associated with higher mean and final performances. %As teams change their composition more frequently, both mean and final performances increase steadily. 
Depending on the scenario studied, each decrease in $\tau$ generally increases mean performance by approximately $2$ to $3$ percentage points and final performance by approximately $3$ to $4$ percentage points. These differences, although small, are statistically significant at a 0.01\% confidence level. Hence, our results are closer to the modern view of dynamic team composition \citep{Bell2017,Tannenbaum2012} than to the traditional perspective on the matter \citep{Carley1992,Carley1996,Davis1973,Lin2006}. Specifically, these results suggest that dynamic team composition is not a negative consequence of suboptimal design choices, as we find no negative effects of increasing the frequency of team formation. Instead, our findings support the characterization of dynamic team composition as a design element consciously implemented by organizations \citep{Tannenbaum2012}. Team composition becomes more efficient when it is periodically reorganized, as \cite{Bell2007} and \cite{Mathieu2014} argue. The mechanism behind this improved efficiency lies in the increased exploration of the solution space attributed to dynamic team composition \citep{March1991,Simon1991}. As prior research argues, this enhaced exploration increases the number of solutions available to the team, thus resulting in the team being more innovative \citep{Choi2004,Choi2005,Hirst2009,Sergeeva2018,Spanuth2020}. Our results show that this increased innovation improves both mean and long-term performance. From a practical standpoint, these findings suggest that organizations should consider incorporating dynamic team composition in their design. As a design element that promotes exploration, this guidance may be particularly relevant for organizations seeking to generate new value, establish a competitive advantage in a particular area, or modernize their structure \citep{Rojas-Cordova2022}.

Regardless of the frequency of team formation, our results suggest that the highest mean and final performances are associated with the lateral communication mode, followed by the sequential coordination mode, the fully autonomous team, and, finally, the liaison coordination mode (see Tab. \ref{table:generaleffects}). Differences in performance between the lateral communication, sequential and fully autonomous modes are quite small, ranging approximately from 1 to 2 percentage points for mean performance and from 3 to 5 percentage points for final performance. All these differences are significant at a 0.01\% confidence level. Conversely, there are significantly larger performance gaps between the lateral communication and liaison coordination modes. Specifically, this gap is of approximately 9 percentage points for mean performance and 12 percentage points for final performance.

These findings follow prior insights into coordination and task performance \citep{Burton2018,Rivkin2003,Siggelkow2005}. In particular, previous results suggest that a proper coordination mode should achieve a balance between autonomy with stability in decision-making \citep{Burton2018,Rivkin2003}. Autonomous decision-making allows team members to intensively explore their available solution space \citep{Siggelkow2005}, and reflects the demand for rapid change and decision-making often encountered in organizations \citep{Nadler1997}. In contrast, stability redirects the team members' focus toward exploiting their available solutions \citep{Siggelkow2005}, which allows them to fully take advantage of their knowledge \citep{Rojas-Cordova2022}. Excessive autonomy, however, might result in overexploration \citep{Rojas-Cordova2022}, reducing task performance in the long run \citep{Siggelkow2005}. This is exemplified in our results by the fully autonomous team. Conversely, excessive stability might slow down decision-making \citep{Nadler1997}, which also decreases task performance \citep{Siggelkow2005}. Our results show that this occurs in the case of the liaison coordination mode. Conversely, the lateral communication and sequential coordination modes balance autonomy and stability in decision-making. The lateral communication mode, for instance, puts more weight on stability but retains a certain degree of autonomous decision-making in the form of the veto power (see Sec. \ref{sec:decision}). The sequential coordination mode, in contrast, focuses more on autonomy but retains some stability in the form of successive decision-making. This balance explains why for these two coordination modes, performance is generally higher than in the fully autonomous team. Consequently, our results suggest that organizations should consider how their coordination choices balance autonomy and stability in decision-making before taking any action.

Additionally, our results suggest that coordination somewhat helps teams exploit the potential positive effects of dynamic composition on task performance. We can observe this by comparing the mean and final performances between a long- and a short-term team composition in Tab. \ref{table:generaleffects}. Differences in mean and final performances between a short- and a long-term team composition are slightly higher for a sequential and a lateral communication mode, approximately amounting to 6 percentage points, than for the fully autonomous team, with around 5 percentage points. This implies that the positive effects of the lateral communication and sequential coordination modes on task performance increase modestly with the frequency of team formation. %Furthermore, differences in final performance are also slightly higher for any of the three coordination modes, around 6 percentage points, than for the fully autonomous team, approximately 5 percentage points. 
Rather than dynamic team composition reducing coordination, as prior research often suggests \citep {Edmondson2001,Edmondsonetal2003,Gardner2010,Harrison2003,Mortensen2014,Reagans2005}, our results indicate a positive---though modest---correlation between these two design elements. %In particular, our findings suggest that coordination slightly reinforces the positive effects of dynamic team composition on task performance. 
Thus, our results are partly in line with those provided by \cite{Spanuth2020}, suggesting an alternative interpretation of the relationship between dynamic team composition and coordination that departs from the traditional perspective on the matter. From a practical standpoint, our results indicate that organizations should ensure that team members effectively coordinate their actions when changes in team composition are frequent, even if the gains in performance are modest.

\subsection{Moderating effects of task complexity}\label{subsec:complex}

\begin{table}[ht]
\caption{Differentiated effects - Low complexity}
\label{table:lowcomplexityeffects}
\renewcommand{\arraystretch}{1.3}
\resizebox{\textwidth}{!}{%
\begin{tabular}{llcccccc}
                             &                             & Block           & Centralized     & Dependent       & Hierarchical    & Local           & Random          \\ \hline
\multirow{8}{*}{Long-term}   & \multirow{2}{*}{Fully autonomous}  & \textbf{0.9344} & 0.8813          & 0.8573          & \textbf{0.8591} & 0.8431          & 0.7970          \\
                             &                             & \textit{($\pm$0.0029)} & \textit{($\pm$0.0036)} & \textit{($\pm$0.0019)} & \textit{($\pm$0.0031)} & \textit{($\pm$0.0026)} & \textit{($\pm$0.0029)} \\
                             & \multirow{2}{*}{Sequential} & \textbf{0.9343} & \textbf{0.9156} & 0.8548          & \textbf{0.8650} & \textbf{0.8483} & 0.8089          \\
                             &                             & \textit{($\pm$0.0029)} & \textit{($\pm$0.0035)} & \textit{($\pm$0.0018)} & \textit{($\pm$0.0029)} & \textit{($\pm$0.0027)} & \textit{($\pm$0.0027)} \\
                             & \multirow{2}{*}{Liaison}    & 0.8427          & 0.7736          & 0.8134          & 0.7698          & 0.7765          & 0.7474          \\
                             &                             & \textit{($\pm$0.0028)} & \textit{($\pm$0.0017)} & \textit{($\pm$0.0020)} & \textit{($\pm$0.0016)} & \textit{($\pm$0.0016)} & \textit{($\pm$0.0019)} \\
                             & \multirow{2}{*}{Lateral}    & 0.9156          & 0.8706          & \textbf{0.8877} & \textbf{0.8656} & \textbf{0.8522} & \textbf{0.8322} \\
                             &                             & \textit{($\pm$0.0025)} & \textit{($\pm$0.0038)} & \textit{($\pm$0.0028)} & \textit{($\pm$0.0038)} & \textit{($\pm$0.0027)} & \textit{($\pm$0.0045)} \\ \hline
\multirow{8}{*}{Medium-term} & \multirow{2}{*}{Fully autonomous}  & \textbf{0.9572} & \textbf{0.9104} & 0.8876          & 0.8858          & 0.8720          & 0.8187          \\
                             &                             & \textit{($\pm$0.0034)} & \textit{($\pm$0.0035)} & \textit{($\pm$0.0024)} & \textit{($\pm$0.0030)} & \textit{($\pm$0.0030)} & \textit{($\pm$0.0027)} \\
                             & \multirow{2}{*}{Sequential} & \textbf{0.9571} & \textbf{0.9156} & 0.8863          & \textbf{0.8928} & \textbf{0.8806} & 0.8361          \\
                             &                             & \textit{($\pm$0.0034)} & \textit{($\pm$0.0034)} & \textit{($\pm$0.0025)} & \textit{($\pm$0.0029)} & \textit{($\pm$0.0031)} & \textit{($\pm$0.0028)} \\
                             & \multirow{2}{*}{Liaison}    & 0.8822          & 0.8110          & 0.8448          & 0.7988          & 0.8059          & 0.7664          \\
                             &                             & \textit{($\pm$0.0030)} & \textit{($\pm$0.0029)} & \textit{($\pm$0.0027)} & \textit{($\pm$0.0026)} & \textit{($\pm$0.0026)} & \textit{($\pm$0.0023)} \\
                             & \multirow{2}{*}{Lateral}    & 0.9435          & 0.9039          & \textbf{0.9166} & \textbf{0.8960} & \textbf{0.8843} & \textbf{0.8576} \\
                             &                             & \textit{($\pm$0.0033)} & \textit{($\pm$0.0042)} & \textit{($\pm$0.0035)} & \textit{($\pm$0.0042)} & \textit{($\pm$0.0042)} & \textit{($\pm$0.0046)} \\ \hline
\multirow{8}{*}{Short-term}  & \multirow{2}{*}{Fully autonomous}  & \textbf{0.9694} & \textbf{0.9327} & 0.9057          & 0.9087          & 0.9010          & 0.8502          \\
                             &                             & \textit{($\pm$0.0033)} & \textit{($\pm$0.0031)} & \textit{($\pm$0.0025)} & \textit{($\pm$0.0027)} & \textit{($\pm$0.0029)} & \textit{($\pm$0.0026)} \\
                             & \multirow{2}{*}{Sequential} & \textbf{0.9695} & \textbf{0.9375} & 0.9059          & 0.9155          & \textbf{0.9123} & 0.8732          \\
                             &                             & \textit{($\pm$0.0033)} & \textit{($\pm$0.0030)} & \textit{($\pm$0.0025)} & \textit{($\pm$0.0027)} & \textit{($\pm$0.0030)} & \textit{($\pm$0.0027)} \\
                             & \multirow{2}{*}{Liaison}    & 0.9266          & 0.8570          & 0.8810          & 0.8396          & 0.8413          & 0.7893          \\
                             &                             & \textit{($\pm$0.0032)} & \textit{($\pm$0.0028)} & \textit{($\pm$0.0029)} & \textit{($\pm$0.0025)} & \textit{($\pm$0.0028)} & \textit{($\pm$0.0023)} \\
                             & \multirow{2}{*}{Lateral}    & \textbf{0.9631} & \textbf{0.9334} & \textbf{0.9419} & \textbf{0.9266} & \textbf{0.9189} & \textbf{0.8910} \\
                             &                             & \textit{($\pm$0.0035)} & \textit{($\pm$0.0040)} & \textit{($\pm$0.0037)} & \textit{($\pm$0.0040)} & \textit{($\pm$0.0043)} & \textit{($\pm$0.0045)} \\ \hline
\end{tabular}}
\end{table}

\begin{table}[ht]
\caption{Differentiated effects - Medium complexity}
\label{table:mediumcomplexityeffects}
\renewcommand{\arraystretch}{1.5}
\resizebox{\textwidth}{!}{%
\begin{tabular}{llcccccc}
                             &                             & Block           & Centralized     & Dependent       & Hierarchical    & Local           & Random          \\ \hline
\multirow{8}{*}{Long-term}   & \multirow{2}{*}{Fully autonomous}  & 0.8150          & 0.8191          & 0.7817          & 0.8254          & 0.7783          & 0.7638          \\
                             &                             & \textit{($\pm$0.0037)} & \textit{($\pm$0.0040)} & \textit{($\pm$0.0027)} & \textit{($\pm$0.0036)} & \textit{($\pm$0.0032)} & \textit{($\pm$0.0031)} \\
                             & \multirow{2}{*}{Sequential} & \textbf{0.8283}        & \textbf{0.8332} & 0.7931          & \textbf{0.8369} & 0.7965          & 0.7835          \\
                             &                             & \textit{($\pm$0.0033)} & \textit{($\pm$0.0035)} & \textit{($\pm$0.0024)} & \textit{($\pm$0.0033)} & \textit{($\pm$0.0028)} & \textit{($\pm$0.0029)} \\
                             & \multirow{2}{*}{Liaison}    & 0.7490          & 0.7233          & 0.7422          & 0.7266          & 0.7319          & 0.7249          \\
                             &                             & \textit{($\pm$0.0022)} & \textit{($\pm$0.0015)} & \textit{($\pm$0.0022)} & \textit{($\pm$0.0014)} & \textit{($\pm$0.0022)} & \textit{($\pm$0.0021)} \\
                             & \multirow{2}{*}{Lateral}    & \textbf{0.8348} & \textbf{0.8391} & \textbf{0.8326} & \textbf{0.8398} & \textbf{0.8182} & \textbf{0.8163} \\
                             &                             & \textit{($\pm$0.0049)} & \textit{($\pm$0.0050)} & \textit{($\pm$0.0050)} & \textit{($\pm$0.0046)} & \textit{($\pm$0.0049)} & \textit{($\pm$0.0051)} \\ \hline
\multirow{8}{*}{Medium-term} & \multirow{2}{*}{Fully autonomous}  & 0.8384          & 0.8460          & 0.8030          & 0.8504          & 0.7950          & 0.7723          \\
                             &                             & \textit{($\pm$0.0034)} & \textit{($\pm$0.0038)} & \textit{($\pm$0.0025)} & \textit{($\pm$0.0033)} & \textit{($\pm$0.0027)} & \textit{($\pm$0.0025)} \\
                             & \multirow{2}{*}{Sequential} & \textbf{0.8576} & \textbf{0.8634} & 0.8205          & \textbf{0.8661} & 0.8225          & 0.8026          \\
                             &                             & \textit{($\pm$0.0033)} & \textit{($\pm$0.0027)} & \textit{($\pm$0.0026)} & \textit{($\pm$0.0033)} & \textit{($\pm$0.0029)} & \textit{($\pm$0.0027)} \\
                             & \multirow{2}{*}{Liaison}    & 0.7660          & 0.7392          & 0.7560          & 0.7496          & 0.7471          & 0.7344          \\
                             &                             & \textit{($\pm$0.0023)} & \textit{($\pm$0.0019)} & \textit{($\pm$0.0022)} & \textit{($\pm$0.0022)} & \textit{($\pm$0.0022)} & \textit{($\pm$0.0020)} \\
                             & \multirow{2}{*}{Lateral}    & \textbf{0.8618} & \textbf{0.8657} & \textbf{0.8558} & \textbf{0.8668} & \textbf{0.8419} & \textbf{0.8348} \\
                             &                             & \textit{($\pm$0.0049)} & \textit{($\pm$0.0050)} & \textit{($\pm$0.0048)} & \textit{($\pm$0.0046)} & \textit{($\pm$0.0048)} & \textit{($\pm$0.0048)} \\ \hline
\multirow{8}{*}{Short-term}  & \multirow{2}{*}{Fully autonomous}  & 0.8698          & 0.8800          & 0.8326          & 0.8800          & 0.8307          & 0.7895          \\
                             &                             & \textit{($\pm$0.0032)} & \textit{($\pm$0.0014)} & \textit{($\pm$0.0024)} & \textit{($\pm$0.0031)} & \textit{($\pm$0.0028)} & \textit{($\pm$0.0025)} \\
                             & \multirow{2}{*}{Sequential} & \textbf{0.8937} & \textbf{0.9012} & 0.8585          & \textbf{0.8979} & \textbf{0.8676} & 0.8457          \\
                             &                             & \textit{($\pm$0.0031)} & \textit{($\pm$0.0032)} & \textit{($\pm$0.0025)} & \textit{($\pm$0.0030)} & \textit{($\pm$0.0029)} & \textit{($\pm$0.0027)} \\
                             & \multirow{2}{*}{Liaison}    & 0.7846          & 0.7617          & 0.7720          & 0.7794          & 0.7648          & 0.7462          \\
                             &                             & \textit{($\pm$0.0023)} & \textit{($\pm$0.0022)} & \textit{($\pm$0.0022)} & \textit{($\pm$0.0023)} & \textit{($\pm$0.0023)} & \textit{($\pm$0.0018)} \\
                             & \multirow{2}{*}{Lateral}    & \textbf{0.8941} & \textbf{0.8953} & \textbf{0.8870} & \textbf{0.9010} & \textbf{0.8746} & \textbf{0.8631} \\
                             &                             & \textit{($\pm$0.0047)} & \textit{($\pm$0.0047)} & \textit{($\pm$0.0047)} & \textit{($\pm$0.0045)} & \textit{($\pm$0.0048)} & \textit{($\pm$0.0047)} \\ \hline
\end{tabular}}
\end{table}

In general, our results suggest that both task complexity and the interdependence structure influence the choice of a proper coordination mode, similarly to the results reported in \cite{Siggelkow2005}. When subtasks are not interdependent (for low complexity and a block diagonal structure, see Fig. \ref{fig:matrix}, Block diagonal (A)), neither of the three coordination modes improves task performance significantly (see the first column of Tab. \ref{table:lowcomplexityeffects}). %For a long-term composition, for example, the mean performance of fully autonomous teams is $0.9344$, against $0.9343$ for teams with sequential decision-making, $0.9156$ for teams with lateral communication, and $0.8427$ for teams with liaison devices.
We report similar results for tasks of low complexity, centralized structure, and teams with a medium- and short-term composition (see the second column of Tab. \ref{table:lowcomplexityeffects}); and for low complexity, hierarchical structure in teams with a long-term composition (see the fourth column of Tab. \ref{table:lowcomplexityeffects}). When interdependencies are concentrated in one subtask (see Fig. \ref{fig:matrix}, Centralized (A) and Hierarchical (A)), the demand for coordination with the remaining team members decreases, as they have no control of the interdependencies.

As interdependencies spread throughout the task, the demand for coordination increases, as reported in prior research \citep{Siggelkow2005}. Specifically, employing either lateral communication or sequential decision-making becomes more advantageous in terms of task performance. In columns three to six of Tab. \ref{table:lowcomplexityeffects} we observe that coordination by sequential decision-making or lateral communication slightly improves task performance in 11 out of the 12 structures. Differences range between approximately 0.5 percentage points for the local interdependence structure and 5 percentage points for the random structure, and increase modestly with the frequency of changing composition. These differences are significant at a 0.01\% confidence level. Among these 11 scenarios, the lateral communication mode is the sole best performer in seven of them. This result is reinforced if we look at tasks of medium complexity, in which there are always interdependencies between subtasks (see Fig. \ref{fig:matrix}). Regarding the 18 scenarios of Tab. \ref{table:mediumcomplexityeffects}, the lateral communication mode shares the highest performer status with the sequential mode in 10 out of them and is the sole highest performer in the remaining eight. Overall, performance increases approximately from 1 percentage point for the hierarchical interdependence structure to 7.5 percentage points for the random interdependence structure. Furthermore, these improvements in task performance are slightly higher for more frequent changes in team composition, and significant at a 0.01\% confidence level. Overall, our findings suggest that lateral communication is the best option for improving task performance when there are interdependencies between subtasks, although in some specific cases sequential decision-making is also an efficient choice.

Prior findings suggest that the interdependence structure has important implications for task performance, as it shapes the performance landscape \citep{Rivkin2007}. We build on these insights by showing that the interdependence structure also affects the choice of the coordination mode. To explain this result, we shall go back to the balance between stability and autonomy in decision-making \citep{Rivkin2003}. As outlined in Sec. \ref{sec:background}, as interdependencies spread throughout the task, uncertainty in decision-making increases, which in turn harms task performance \citep{Levinthal1997,Lin1992,Wall2016,Wall2018}. Decision-makers need to counteract this uncertainty by making design choices that ensure stability, i.e., by focusing on coordination choices \citep{Galbraith1973,Siggelkow2005}. Among these, lateral communication is usually the most efficient, although sequential decision-making may also be an optimal choice in certain cases. From a practical perspective, our results imply that organizations should carefully review the nature of the tasks they face and evaluate how different coordination modes may fit their demands. 

These differentiated results also reinforce the insights discussed in Sec. \ref{subsec:general} regarding the interrelations between coordination and dynamic team composition. Specifically, while both task complexity and the interdependence structure moderate the demand for coordination, they do not affect the fundamental relationship between coordination and dynamic team composition. Again, to illustrate this we examine the performances for the fully autonomous team and compare it to the three coordination modes, both for a long- and a short-term composition. % In tasks of low complexity, we see that the difference in performance between a short- and a long-term team composition is slightly higher for the lateral communication mode in four interdependence structures (Block diagonal, Centralized, Hierarchical, and Local), for the liaison mode in four interdependence structures (Block diagonal, Centralized, Dependent, and Hierarchical), and for the sequential decision-making mode in one interdependence structure (Random). In tasks of moderate complexity, the difference in performance is higher for the sequential mode in all structures, while differences are higher for the lateral communication mode in two interdependence structures (Hierarchical and Random). 
Results show that for all kinds of tasks considered teams that follow a specific coordination mode gain slightly more from changing their composition periodically than fully autonomous teams. Thus, the results of Sec. \ref{subsec:general} regarding the interrelations between coordination and dynamic team composition are robust to variations in task complexity and the interdependence structure.

\subsection{Moderating effects of individual learning}\label{subsec:learning}
%Our findings suggest that learning has different effects depending on team composition and the coordination mode.

%
\begin{figure}[ht]
 \centering
 \includegraphics[width=\textwidth]{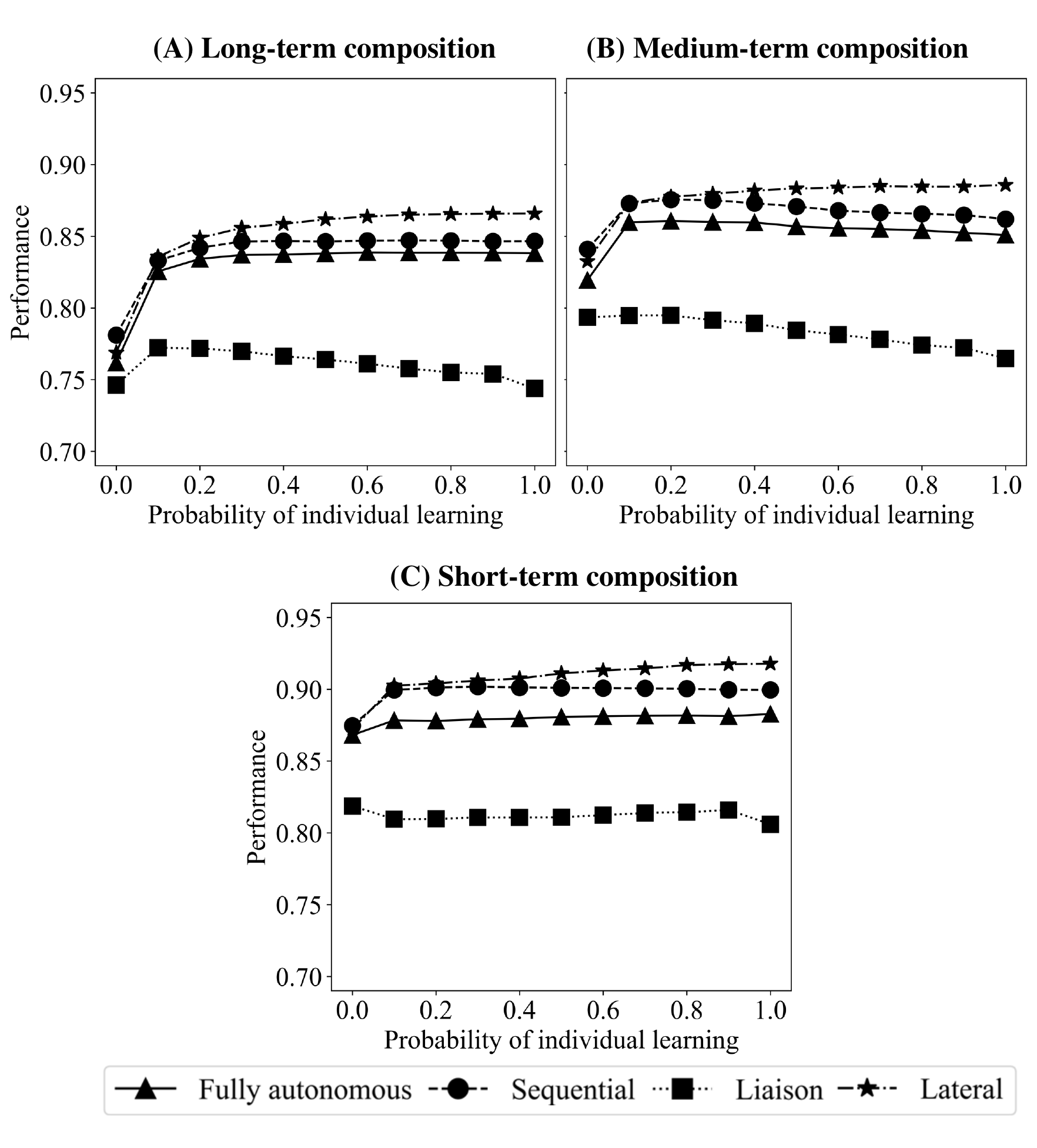}
 \caption{Mean performance for each probability of learning and team composition.}
 \label{fig:learningeffects}
\end{figure}

In Fig. \ref{fig:learningeffects}, we plot the mean performances achieved by each team depending on the coordination mode and the learning probability, i.e., $\mathbb{P}$. Additionally, we differentiate the results by team composition. In general, increasing the learning probability has more significant effects the less frequently teams change their composition. For example, task performance in a fully autonomous team with a long-term composition increases from 0.7617 at $\mathbb{P}=0$ to 0.8381 at $\mathbb{P}=1$ (see Fig. \ref{fig:learningeffects}(A)). When the fully autonomous team has a short-term composition, however, performance increases from 0.8680 only to 0.8827 (see Fig. \ref{fig:learningeffects}(C)). This pattern is robust to all coordination modes and consistent with prior results \citep{Blanco-Fernandez2023,Blanco2023}. 

For the lateral communication mode, the sequential coordination mode, and the fully autonomous team, Fig. \ref{fig:learningeffects} shows how performance increases significantly when agents start to learn, i.e., when $\mathbb{P}$ increases from 0 to $0.1$. For instance, in the lateral communication mode, performance increases by approximately 6.5, 4, and 3 percentage points for a long-, medium-, and short-term composition, respectively. These differences in performance are statistically significant at a 0.01\% confidence level. An increase in the learning probability beyond $0.1$ improves task performance with decreasing marginal effects. Once the learning probability reaches a certain threshold, performance slightly falls if team composition is dynamic for the sequential coordination mode and the fully autonomous team (see Fig. \ref{fig:learningeffects}(B) and (C)). The cumulative negative effects are around 1 percentage point and lack statistical significance at a 0.01\% confidence level. Although these negative effects are not significant, it is worth noting that they do not appear in the lateral communication mode. Instead, we can observe a slightly steady increase in performance---approximately 1 percentage point in total---as the learning probability increases above $\mathbb{P}=0.1$.

Conversely, we can observe a different relationship between individual learning and task performance in the liaison coordination mode. For a long-term team composition, there is an increase in task performance when agents start learning (see Fig. \ref{fig:learningeffects}(A)). As the learning probability increases further, however, performance declines steadily. In total, performance declines around 3 percentage points between $\mathbb{P}=0.1$ and $\mathbb{P}=1$, which is statistically significant at a 0.01\% level. This steady decline in task performance is also observable for a medium-term composition (see Fig. \ref{fig:learningeffects}(B)), also adding up to approximately 3 percentage points. For a short-term team composition (see Fig. \ref{fig:learningeffects}(C)), we observe a slight decrease in performance when agents start to learn. The performance then stabilizes as the learning probability increases. Finally, performance declines once the probability reaches $\mathbb{P}=1$. Overall, performance declines by approximately 1.25 percentage points, which is statistically significant at a 0.01\% level. 

Our findings show that coordination choices influence the interrelations between individual learning and dynamic team composition, and their effects on task performance. The impact of learning is lower in dynamic teams because, as \cite{Simon1991} notes, dynamic team composition and individual learning are exploratory processes with similar effects. Specifically, they are both mechanisms that increase the number of solutions available to the team. Thus, if one is already present, the marginal effect on performance of the other is reduced \citep{Blanco-Fernandez2023,Blanco2023}. Furthermore, excessive individual learning may result in decreases in task performance. As \cite{Rojas-Cordova2022} note, overexploration may result in long-term problems for organizations. For instance, financial resources may be reduced if innovation is not associated with significant improvements in performance \citep{Rojas-Cordova2022}. From a practical perspective, this means that organizations should consider the balance among the different exploration processes that take place---i.e., individual learning and dynamic team composition---as our results show that this is a highly relevant issue. Organizations in which dynamic team composition is prevalent may not gain from investing in individual learning, and vice versa.

To prevent this overexploration, prior research suggests various methods to shift resources towards the exploitation of current knowledge \citep{Hakonsson2016,Rojas-Cordova2022}. Among all methods, prior research emphasizes the establishment of coordination among decision-makers as one of the most relevant methods \citep{Hakonsson2016,Rojas-Cordova2022}. Our results are consistent with these insights, as they show that teams \textit{(i)} employing either lateral communication or sequential decision-making and \textit{(ii)} maintaining a long-term team composition experience greater advantages from individual learning. Coordination, thus, may enable teams to exploit the beneficial effects of individual learning on task performance. This exploitation is particularly relevant for stable teams, although dynamic teams may also benefit from implementing coordination (see Fig. \ref{fig:learningeffects}(C)). Not all coordination choices, however, have the same effects. Suboptimal coordination choices can decrease task performance, as results for the liaison coordination mode suggest. Conversely, lateral communication eliminates the decline in task performance associated with excessive learning, even when teams change their composition very frequently (see Fig. \ref{fig:learningeffects}(A)-(C)).

\section{Conclusion}\label{sec:conclusion}
In this paper, we employ an agent-based modeling approach to study the interrelations of dynamic team composition, coordination, and task performance. Furthermore, we investigate the moderating role of task complexity and individual learning. Our results depart from the traditional perspective on dynamic team composition, which interprets it as a result of suboptimal design choices that reduce coordination and decrease task performance. Instead, we show that dynamic team composition may be characterized as a design element that improves the innovative capacity of teams. According to our results, this improves task performance. 

Characterizing dynamic team composition as a design element enables us to study how it interacts with another design element, namely the coordination mode. Departing from the traditional perspective on dynamic team composition allows us to find new insights regarding these elements and their interplay. According to our results, coordination helps teams to exploit the positive effects of dynamic team composition on task performance. This is particularly relevant for complex tasks, as coordination reduces the uncertainty in decision-making associated with complexity. Specifically, we find that both lateral communication and sequential decision-making improve task performance, while liaison devices reduce it.

Based on our results, we can argue that organizations increasingly rely on dynamic team composition because it allows teams to adapt to the task environment over time, improving task performance in the process. Thus, we can study how dynamic team composition interrelates with other adaptation processes, such as individual learning. Our results suggest that combining dynamic team composition with individual learning may lead to overexploration, and performance may not increase significantly. To reduce these effects, organizations may resort to coordination choices. For instance, our results indicate that lateral communication may successfully prevent the negative effects of excessive exploration.
 
Nevertheless, our research is not without limitations. First, we work under the implicit assumption of zero coordination costs \citep{Siggelkow2005,Wall2018}. Future research could consider them. Second, we do not study the potential effects of changing the agents' incentives and the interrelations between these incentives and the coordination mode \citep{Siggelkow2005}. Finally, we employ a classic approach to organizational design, taking the design choices as given and studying their effects over time. Further extensions of our research could employ an evolutionary perspective to organizational design in which design choices emerge over time \citep{Leitner2024,Wall2018}.

\backmatter

\section*{Declarations}
\subsection*{Funding}
Does not apply.

\subsection*{Conflict of interest/competing interests}
The authors declare that they have no conflict of interest.

\subsection*{Availability of data}
Simulation data is available  \href{here.}{https://figshare.com/articles/dataset/Datasets_-_RMSC/24598278}

\subsection*{Code availability}
The code is available \href{here.}{https://figshare.com/articles/software/Source_code/24598290}

\newpage
\bibliography{sn-bibliography.bib}% common bib file
\newpage
%\newpage
\begin{appendices}

\section{Technical aspects of the model}\label{app:technical}
\subsection{Task environment and performance landscape}\label{app:landscape}
We model a complex task as an $N$-dimensional binary vector such that $\mathbf{d}=\left(d_1, \dots, d_N \right)$. Each decision $d_n$ contributes $c_n$ to the overall task performance $C(\mathbf{d})$. Task performance is the average of all contributions, such that:

\begin{equation}
    \label{eq:taskperf}
    C(\mathbf{d})=\frac{1}{N}\sum_{n=1}^{N}c_{n}.
\end{equation}

\noindent There are $2^N$ different solutions to the complex task, each with an associated performance. The mapping of each of the $2^N$ solutions to its associated performance is the \textit{performance landscape}. The performance landscape can be conceptualized as a multidimensional space in which the $N$ decisions are represented in the horizontal axis and their associated performances in the vertical axis \citep{Rivkin2007}. The concept of performance landscape is analogous to that of the \textit{fitness landscape} of evolutionary biology \citep{Kauffman1993}. In its original formulation, \cite{Kauffman1993} sought to explain the population dynamics of a certain organism and the fitness of its characteristics to a particular environment. In this formulation, an organism is formed by $N$ interdependent genes that evolve over time. Periodically, the organisms below a certain fitness threshold are assumed to have not survived the environmental conditions. As an agent-based modeling approach, the original formulation of the NK framework examines how the different behavior of multiple individuals---in this case, assumed to be organisms formed by a set of genes---lead to emergent properties at the collective level---in this case, the population dynamics of the organism in a particular environment \citep{Kauffman1993}. \cite{Levinthal1997} adapted the idea of the NK framework and the fitness landscapes to explain the population dynamics of organizations and their evolution based on each organization's initial conditions, whereby $N$ represents the different design elements of an organization rather than genes. Thus, the overall fitness achieved by an organization determines its likelihood of survival. Similarly to its biological counterpart, \cite{Levinthal1997} examines how the different design choices of each individual organization eventually lead up to alterations in their population.

Further adaptations of the NK framework have abandoned this emphasis on population dynamics and focus on aspects such as departmental decision-making \citep{Siggelkow2005,Wall2018}, incentive systems \citep{Leitner2021}, and decision-making in teams \citep{Billinger2014,Blanco-Fernandez2023,Blanco2023,Giannoccaro2019}. Rather than modeling and observing the behavior of large organizations that operate in an environment, the scope is reduced to examine human decision-makers---the agents of the model---that operate in a team or in an organization and face a complex task. In these adaptations, the term \textit{fitness landscape} is often replaced by the term \textit{performance landscape} to emphasize that their primary focus is not on population dynamics. We follow these further adaptations, characterizing $N$ as the interdependent decisions of a task that have an associated task performance.

The core concept underlying the NK framework is the investigation of adaptive search processeses \citep{Wall2016}. This implies that studies employing the NK framework normally model a set of agents who engage in a \textit{neighborhood search} process on the performance landscape, looking for new solutions to the complex task that improve task performance \citep{Levinthal1997,Rivkin2007}. The neighborhood search process works as follows \citep{Levinthal1997}. Agents randomly change the value of one of the $N$ decisions and observe its associated performance. This new decision is adopted if it improves task performance compared to the previous period. Conversely, this decision is rejected if performance declines. As this development continues, the agents follow a \textit{hill-climbing} process on the performance landscape. This process stops when none of the immediate alternatives improve task performance, i.e., when agents reach a peak. Depending on the shape of the performance landscape, this peak may be global---i.e., the optimal decision---or local---i.e., a suboptimal decision.

The shape of the performance landscape is influenced by task complexity \citep{Levinthal1997,Rivkin2007}. Each contribution $c_n$ depends on its associated decision $d_n$ and, due to the existence of interdependencies, on $K$ other decisions. This means that $c_{n}=f(d_{n}, d_{i_1}, \dots, d_{i_K})$, where $\{i_1, \dots, i_K \} \subseteq \{1, \dots, n-1, n+1, \dots, N \}$ and $0 \leq K \leq N-1$. If there are no interdependencies, i.e., $K=0$, the performance landscape is single-peaked. The landscape becomes more rugged as $K$ increases and approaches $N-1$ \citep{Levinthal1997,Rivkin2007}. Rugged landscapes are characterized by several local optima where agents might get stuck \citep{Levinthal1997,Rivkin2007}. This characterization of complexity is a crucial feature of the NK framework. Specifically, the NK framework allows for the the control and manipulation of task complexity \citep{Wall2016}. This, in turn, allows the modeler to observe the changes in behavior of individual agents that result from altering $K$---and, thus, the shape of the performance landscape.

Furthermore, \cite{Rivkin2007} argue that not only the number of interdependencies---i.e., $K$---influences the shape of the performance landscape, but also the \textit{interdependence structure}. The interdependence structure is reflected in the interdependence matrix, which is a $N \times N$ matrix that represents contributions in one axis and decisions on the other. Interdependencies are usually marked with an \textit{x} (see Fig. \ref{fig:matrix} in Sec. \ref{sec:variables}). The interdepedence structure often varies depending on the task's nature \citep{Rivkin2007}. As prior results show, the number of local peaks decreases when interdependencies concentrate on few decisions, making the global peak more accessible for agents \citep{Blanco-Fernandez2023,Rivkin2007}.

\subsection{Agents}\label{app:agents}

Approaches based on the NK framework usually assume that agents have limited capabilities and are thus rationally bounded \citep{Wall2016,Wall2020}. Specifically, it is often assumed that the agents' information processing-power and their access to it are limited \citep{Simon1957}. This assumption implies that, rather than looking immediately for an optimal solution, agents search locally and step-wise on the performance landscape \citep{Wall2016,Wall2020}. Furthermore, it is often assumed that agents are heterogeneous in one or more aspects \citep{Wall2016,Wall2020}. To reflect the limited capabilities of agents and their heterogeneity, we partition the $N$-dimensional task into $M$ subtasks of equal size $S=N/M$ and randomly assign each agent to one of them. Each subtask is denoted by $\mathbf{d}_m = (d_{S\cdot(m-1)+1},\dots, d_{S\cdot m})$. Additionally, agents do not start with full knowledge of the solution space. Instead, we endow each agent with one random solution to their assigned subtask $\mathbf{d}_m$, and they learn new solutions over time (see Sec. \ref{sec:learning} and Appendix \ref{app:learning}).

The $P$ agents of the model are utility-maximizers and myopic, experiencing utility only when they are part of the team. Specifically, a team member assigned to subtask $m$ maximizes the following utility function:

\begin{equation}\label{eq:utility}
   U(\mathbf{d}_{mt}, \mathbf{D}_{mt}) = \frac{1}{2} \cdot \left( C(\mathbf{d}_{mt}) + \frac{1}{M-1} \sum_{\substack{{r=1}\\{r\neq m}}}^{M} C(\mathbf{d}_{rt})\right) ~,
\end{equation}

\noindent where $C(\mathbf{d}_{mt})$ is the performance contribution coming from the agent \textit{m}'s own area of expertise. Additionally, $C(\mathbf{d}_{rt})$, where $r=\{1,\dots,M\}\subset \mathbb{N}$ and $r \neq m$, is the contribution from the \textit{residual decisions} $\mathbf{D}_{mt} = (\mathbf{d}_{1t}, \dots, \mathbf{d}_{\{m-1\}t}, \mathbf{d}_{\{m+1\}t},\dots, \mathbf{d}_{Mt})$, i.e., the contribution of the decisions located outside subtask $m$.

\subsection{Team formation}\label{app:team}
Agents self-organize in a team to solve the complex task. They follow a signaling mechanism for team formation that works as follows. Agent $m$ knows a set of solutions denoted by ${\mathbf{S}}_{m} = \{\hat{\mathbf{d}}_{m1},\dots,\hat{\mathbf{d}}_{mI}\}$ where $\hat{\mathbf{d}}_{mi}$ is a solution to subtask $\mathbf{d}_m$, $i=\{1,\dots,I\} \subset \mathbb{N}$ and $1 \leq I \leq 2^S$. Each agent estimates the utility associated with each solution they know, i.e., $\forall \hat{\mathbf{d}}_{mi} \in {\mathbf{S}}_{m}$. Since agents cannot predict the residual decisions for the next period, they assume that the residual decisions from the previous period $\mathbf{D}_{m\{t-1\}}$ will not change. Agent $m$'s \textit{estimated utility} is then:

\begin{equation}\label{eq:estutility}
    EU(\mathbf{d}_{mt}, \mathbf{D}_{m\{t-1\}}) = \frac{1}{2} \cdot \left( C(\mathbf{d}_{mt}) + \frac{1}{M-1} \sum_{\substack{{r=1}\\{r\neq m}}}^{M} C(\mathbf{d}_{r\{t-1\}})\right) \cdot \left(1 + e \right) ~;
\end{equation}

\noindent where $e$ is an error term that reflects the miscalculations that agents might make. %Reporting errors and miscalculations are often found in organizational databases \citep{Tee2007}. 
Prior research estimates that, on average, error rates range between 1 and 10 percent \citep{Tee2007}. To reflect this, the error term follows a normal distribution with a standard deviation of $0.1$, i.e., $e\sim N(0,0.1)$. This formulation is consistent with the NK framework. Specifically, \cite{Levinthal1997} assumes in his adaptation of the NK framework that the agents of the model---i.e., the organizations---are able to identify the fitness value of each solution. \cite{Levinthal1997} introduces and examines the concept of \textit{noisy search}, which occurs when agents can commit errors in the identification of the fitness values. We follow a similar approach in our formulation by assuming that the agents are able to estimate the utility each solution reports, although they may commit small errors.

Each agent then sends a signal equivalent to the highest estimated utility available, i.e., their signal is $U(\hat{\mathbf{d}}^{*}_{mt}, \mathbf{D}_{m\{t-1\}})$, where $\hat{\mathbf{d}}^{*}_{mt} := \argmax_{\mathbf{d}^\prime \in \mathbf{S}_{mt}} ~ U(\mathbf{d}^\prime, \mathbf{D}_{m\{t-1\}})$. The agent with the highest signal for subtask $\mathbf{d}_m$ joins the team. Consequently, the team is formed by $M$ agents.

\subsection{Decision-making and coordination}\label{app:coordination}
We model three coordination modes and one benchmark scenario without coordination. In this benchmark scenario we model a fully autonomous team in which decision-making is completely decentralized and team members do not communicate. Each team member estimates the utility for each solution they know, i.e., $\forall \hat{\mathbf{d}}_{mi} \in \mathbf{S}_{mt}$, following Eq. \ref{eq:estutility}. Then, each member proposes the solution to their subtask associated to the highest estimated utility, $\hat{\mathbf{d}}^{*}_{mt}$. The overall task solution for the current period is the concatenation of all members' proposals, denoted by $\mathbf{d}_{t} :=\hat{\mathbf{d}}^{*}_{1t}\frown \dots \frown \hat{\mathbf{d}}^{*}_{Mt}$.

In the sequential coordination mode, the proposals are successively made. The team member associated with subtask $m$ chooses a solution to their subtask and reports it to the $M-m$ following members, who update their residual decisions accordingly. They repeat this process until the $M$ members have made their proposals. The team member associated with the first subtask makes their proposal according to previous period residual decisions $\mathbf{D}_{1\{t-1\}}$, while the other team members make their proposal using their updated residual decisions as a basis for their estimations $\mathbf{D}_{m} = (\mathbf{d}_{1t}, \dots, \mathbf{d}_{\{m-1\}t}, \mathbf{d}_{\{m+1\}\{t-1\}},\dots, \mathbf{d}_{M\{t-1\}})$, where $m \neq 1$. The concatenation of all the sequentially-made proposals forms the overall solution to the complex task, i.e., $\mathbf{d}_{t} :=\hat{\mathbf{d}}^{*}_{1t}\frown \dots \frown \hat{\mathbf{d}}^{*}_{Mt}$.

The liaison coordination mode works as follows: Each member ranks each solution known in terms of their estimated utility. Then, they choose their two highest-ranked proposals, $\hat{\mathbf{d}}^{(1)}_{mt}$ and $\hat{\mathbf{d}}^{(2)}_{mt}$, taking them to a coordination meeting. During this meeting, two candidate solutions are created: the first by concatenating the team members' top-choice proposals and the second by combining their second-choice proposals, such that $\mathbf{d}^{(j)}_{t} :=\hat{\mathbf{d}}^{(j)}_{1t}\frown \dots \frown \hat{\mathbf{d}}^{(j)}_{Mt}$ where $\hat{\mathbf{d}}^{(j)}_{mt}$ is agent $m$'s $j^{th}$ preferred choice. These candidate solutions are evalutated by the team members according to their estimated utility. Team members accept the candidate solution if they estimate that their utility will increase, i.e., if $EU_m(\mathbf{d}^{(j)}_{t} )>U_m(\mathbf{d}_{t-1})$. Otherwise, they veto the candidate solution. In case both candidate solutions are vetoed, the overall solution does not change from the prior period, so $\mathbf{d}_{t}=\mathbf{d}_{t-1}$. Conversely, if one candidate solution is accepted, it is implemented in the upcoming period, such that $\mathbf{d}_{t}=\mathbf{d}^{(j)}_{t}$.

Finally, in the lateral communication mode each member randomly picks two solutions to their assigned subtask from the set of known solutions ${\mathbf{S}}_{mt}$ and brings them to a coordination session, i.e., picking  $\hat{\mathbf{d}}^{(1)}_{mt}, \hat{\mathbf{d}}^{(2)}_{mt} \sim U(\mathbf{S}_{mt})$. Then, two candidate solutions are formed by randomly concatenating the team members' picks, i.e., $\mathbf{d}^{(j)}_{t} :=\hat{\mathbf{d}}_{1t}\frown \dots \frown \hat{\mathbf{d}}_{Mt}$ where $\hat{\mathbf{d}}_{mt} \sim U \left(\hat{\mathbf{d}}^{(1)}_{mt}, \hat{\mathbf{d}}^{(2)}_{mt}\right)$. Then, each team member accepts or vetoes the candidate solutions based on their estimated utility, i.e., depending on whether $EU_m(\mathbf{d}^{(j)}_{t} )>U_m(\mathbf{d}_{t-1})$ or not. The candidate solution is implemented if all members accept it. Conversely, if both candidate solutions are vetoed, the overall solution remains the same as in the previous period.

\subsection{Learning}\label{app:learning}
Following the principles of the NK framework, our agents cannot observe the complete set of solutions to their assigned subtask $\mathbf{d}_m$ from the beginning. Instead, they learn step-wise, adapting their set of known solutions ${\mathbf{S}}_{mt}$ over time. Individual learning consists of two processes: discovering and forgetting. Agent $m$ discovers a solution to their subtask with probability $\mathbb{P}$. This solution is randomly picked from all unknown solutions. The distribution of the probabilities with which the solutions are discovered is based on agent $m$'s beliefs. Agent $m$ updates these beliefs according to a Bayesian updating rule.We denote agent $m$'s belief about solution $\hat{\mathbf{d}}_{mi}$ by:

\begin{equation}\label{eq:learn}
    p^{(i)}_{mt} = \frac{\alpha^{(i)}_{mt}}{\alpha^{(i)}_{mt}+\beta^{(i)}_{mt}}~,
\end{equation}

\noindent where $\alpha^{(i)}_{m1}=\beta^{(i)}_{m1} = 1$ so that $p^{(i)}_{m1}=0.5$. This means that, in the beginning, all solutions have equal beliefs, i.e., that the initial probability distribution with which solutions are discovered is uniform. Beliefs, however, may change as time passes, altering this probability distribution. Each time agent $m$ implements solution $\hat{\mathbf{d}}_{mi}$, they experience utility $U_m(\mathbf{d}_{t})$. Then, they assess this utility and compare it to the utility they experienced in the previous period $U_m(\mathbf{d}_{t-1})$. If solution $\hat{\mathbf{d}}_{mi}$ improves agent $m$'s utility compared to the previous period, $\alpha^{(i)}_{m}$ will increase by one. This, in turn, increases the probability of discovering solution $\hat{\mathbf{d}}_{mi}$. Conversely, $\beta^{(i)}_{m}$ will increase by one if solution $\hat{\mathbf{d}}_{mi}$ decreases agent $m$'s utility compared to the previous period. The probability of discovering solution $\hat{\mathbf{d}}_{mi}$ decreases as a result. We formalize this process in Eqs. \ref{eq:learn_a} and \ref{eq:learn_b}.

\begin{subequations}\label{eq:updatelearn}
    \begin{equation}\label{eq:learn_a}
        \alpha^{(i)}_{mt}
            \begin{cases}
                \alpha^{(i)}_{m\{t-1\}}+1 \textrm{ if } U_m(\mathbf{d}_{t}) \geq U_m(\mathbf{d}_{t-1}),\\
                \alpha^{(i)}_{m\{t-1\}} \textrm{ otherwise.}
            \end{cases}
    \end{equation}
    \begin{equation}\label{eq:learn_b}
        \beta^{(i)}_{mt} 
            \begin{cases}
                \beta^{(i)}_{m\{t-1\}}+1 \textrm{ if } U_m(\mathbf{d}_{t}) < U_m(\mathbf{d}_{t-1}),\\
                \beta^{(i)}_{m\{t-1\}} \textrm{ otherwise.}
            \end{cases}
    \end{equation}
\end{subequations}

\noindent \noindent Thus, a belief $p^{(i)}_{mt}$ is only updated when agent $m$ knows and implements solution $\hat{\mathbf{d}}_{mi}$.

\noindent Agents may forget one known solution with probability $\mathbb{P}$. The forgotten solution is picked randomly from the pool of known solutions ${\mathbf{S}}_{mt}$. The corresponding probability distribution with which solutions are forgotten is based on agent $m$'s beliefs. Specifically, agent $m$'s belief about solution $\hat{\mathbf{d}}_{mi}$ follows:

\begin{equation}\label{eq:forget}
    q^{(i)}_{mt} = \frac{\lambda^{(i)}_{mt}}{\lambda^{(i)}_{mt}+\delta^{(i)}_{mt}}~.
\end{equation}

\noindent Initially, at $t=1$, all solutions have equal beliefs, such that $\lambda^{(i)}_{m1}=\delta^{(i)}_{m1} = 1$ and $q^{(i)}_{m1}=0.5$. This means that agents forget solutions with a uniform probability distribution. This probability distribution, however, may change as time passes and agents update their beliefs. Each time agent $m$ proposes solution $\hat{\mathbf{d}}_{mi}$, they update belief $q^{(i)}_{mt}$ by evaluating the current utility and comparing it to the one experienced in the previous period. $\lambda^{(i)}_{m}$ will decrease by one if solution $\hat{\mathbf{d}}_{mi}$ reduces agent $m$'s utility compared to the previous period. Consequently, the probability of forgetting solution $\hat{\mathbf{d}}_{mi}$ increases. Conversely, if solution $\hat{\mathbf{d}}_{mi}$ improves agent $m$'s utility, $\delta^{(i)}_{m}$ increases by one, and the probability of forgetting solution $\hat{\mathbf{d}}_{mi}$ decreases.

Additionally, we incorporate a memory factor into the agents' updating process, taking into account that agents tend to forget solutions that have been stored in their memory for an extended period. This means that the probability of forgetting each solution increases as time passes, i.e., $\forall \hat{\mathbf{d}}_{mi} \in \mathbf{S}_{mt}: \lambda^{(i)}_{mt} = \lambda^{(i)}_{m\{t-1\}}+1$. 

Thus, if agent $m$ implements solution $\hat{\mathbf{d}}_{mi}$ and their utility decreases, $\lambda^{(i)}_{mt}$ will increase by two. $\lambda^{(i)}_{mt}$ increases by one because one period has passed and, additionally, %$\lambda^{(i)}_{mt}$ further increases 
by one because utility has decreased. This double increase makes it more likely to forget $\hat{\mathbf{d}}_{mi}$. We provide the general updating rule for belief $q^{(i)}_{mt}$ in Eqs. \ref{eq:forget_a} and \ref{eq:forget_b}.

\begin{subequations}\label{eq:updateforget}
    \begin{equation}\label{eq:forget_a}
        \lambda^{(i)}_{mt}
            \begin{cases}
                \lambda^{(i)}_{m\{t-1\}}+2 \textrm{ if } U_m(\mathbf{d}_{t}) < U_m(\mathbf{d}_{t-1}),\\
                \lambda^{(i)}_{m\{t-1\}}+1 \textrm{ otherwise.}
            \end{cases}
    \end{equation}
    \begin{equation}\label{eq:forget_b}
        \delta^{(i)}_{mt}
            \begin{cases}
                \delta^{(i)}_{m\{t-1\}}+1 \textrm{ if } U_m(\mathbf{d}_{t}) > U_m(\mathbf{d}_{t-1}),\\
                \delta^{(i)}_{m\{t-1\}} \textrm{ otherwise.}
            \end{cases}
    \end{equation}
\end{subequations}

\subsection{Performance measures}\label{app:perf}
Our dependent variable is the observed task performance at each period, $C(\mathbf{d_{t\phi}})$, where $\phi \in [0,...,\Phi]$ corresponds to the current simulation round. To compare the different scenarios, we normalize $C(\mathbf{d_{t\phi}})$ by the maximum achievable performance in each simulation round, i.e., $C^{\ast}_{\phi}$. We calculate the mean normalized performance at each period for the 1,500 simulation rounds, $\Bar{C}_t$. This measure is formalized in Eq. \ref{eq:mean-perf}.

\begin{equation}
\label{eq:mean-perf}
    \overline{C}_t = \frac{1}{\Phi} \sum_{\phi=1}^{\Phi} \frac{C\left( \mathbf{d}_{t\phi}\right)}{C^{\ast}_\phi}~.
\end{equation}

In our results (see Sec. \ref{sec:results}), we report two performance measures. The first performance measure corresponds to the mean performance of the $100$ periods of each simulation, and it follows:

\begin{equation}
\label{eq:meanperf}
    \overline{\overline{C}}_t = \frac{1}{T} \sum_{t=1}^{T}\overline{C}_t ~,
\end{equation}

\noindent where $\overline{C}_t$ comes from Eq. \ref{eq:mean-perf} and $T=100$ is the last observable period. The second performance measure is the final performance, which is the performance achieved on average by a team at $t=100$, i.e., $\overline{C}_{100}$.

\subsection{Other technical details}\label{app:details}

We implemented and ran the model in Python 3.7.4. using the Spyder software, version 5.0.0. Two laptops were employed in the simulations: One with 16GB of RAM and a 1.90 GHz Intel Core i7 Processor and the other with 8GB of RAM and a 3.30 GHz AMD Ryzen 5 5600H with Radeon Graphics processor. Each simulation takes between 10 and 40 minutes, depending on the device and the settings.

\end{appendices}




\end{document}